\documentclass{emulateapj}
\usepackage{apjfonts}

\journalinfo{Accepted for publication in the Astrophysical Journal}
\slugcomment{Submitted to ApJ November 9, 2010; accepted March 10, 2011}

\newcommand{\etal}{et al.}
\newcommand{\snia}{SN~Ia}
\newcommand{\sneia}{SNe~Ia}
\newcommand{\snf}{SNfactory}
\newcommand{\superc}{super-Chandrasekhar}
\newcommand{\oii}{[OII]~$\lambda\lambda 3727,3730$}

\newcommand{\oiii}{[OIII]~$\lambda\lambda 4959,5007$}
\newcommand{\oiiia}{[OIII]~$\lambda 4959$}
\newcommand{\oiiib}{[OIII]~$\lambda 5007$}
\newcommand{\aoiii}{[OIII]~$\lambda 4363$}

\newcommand{\hb}{H$\beta$}
\newcommand{\ha}{H$\alpha$}
\newcommand{\nii}{[NII]~$\lambda\lambda 6548,6584$}
\newcommand{\niia}{[NII]~$\lambda 6548$}
\newcommand{\niib}{[NII]~$\lambda 6584$}

\newcommand{\siia}{[SII]~$\lambda 6717$}
\newcommand{\siib}{[SII]~$\lambda 6731$}

\newcommand{\finalappmag}{$m_g=23.15\pm0.06$}
\newcommand{\finalzlines}{$z_\mathrm{lines}=0.074500\pm0.000010$}
\newcommand{\finalzhelio}{$z_{helio}=0.07450\pm0.00015$}
\newcommand{\finalzcmb}{$z_\mathrm{CMB}=0.07336\pm0.00015$}
\newcommand{\finaldistmod}{$\mu=37.60\pm0.004$}
\newcommand{\finalabsmag}{$M_g=-14.45\pm0.06$}
\newcommand{\finalugcolor}{$u-g=0.67\pm0.03$~mag} 
\newcommand{\finalgrcolor}{$g-r=0.07\pm0.04$~mag} 

\newcommand{\finalsfrhalpha}{$SFR_{H\alpha}=2.2\times10^{-3}~M_\odot~yr^{-1}$}
\newcommand{\finalsfrburst}{$SFR_{burst}\approx0.1~M_\odot~yr^{-1}$}
\newcommand{\finalsfrratio}{$SFR_{burst}/SFR_{H\alpha}\approx50$}
\newcommand{\finalmetalkk}{$12+\log(O/H)_\mathrm{KK04}=8.01\pm0.09$}
\newcommand{\finalionpar}{$q=1.46\pm0.48\times10^7$}
\newcommand{\finalmetaltrem}{$12+\log(O/H)_\mathrm{T04}=7.71\pm$0.14[stat]$\pm$0.06[sys]}
\newcommand{\finalhostlogage}{$\log(t)=8.09^{+0.37}_{-0.43}$[stat]$\pm0.06$[sys]}
\newcommand{\finalhostage}{$t_\mathrm{burst}=123^{+165}_{-77}$~Myr}
\newcommand{\finalageapprox}{123~Myr}
\newcommand{\finalhostmsto}{$M/M_\odot=4.6^{+2.6}_{-1.4}$}
\newcommand{\finalmstoplain}{$M/M_\odot=4.6$}
\newcommand{\finalwdmass}{$M_{WD}=0.85~M_\odot$}
\newcommand{\finalsystemmass}{$M_{tot}=1.70~M_\odot$}
\newcommand{\solarglum}{$M_{\odot,g}=5.15$}
\newcommand{\finalmtolgsolar}{$\log(M_*/L)_\mathrm{model}=-0.50\pm0.17$}
\newcommand{\finalhostmass}{$\log(M_*/M_\odot)=7.32\pm0.17$}
\newcommand{\finalmtolgsdss}{$\log(M_*/L)_\mathrm{SDSS}=-0.52\pm0.15$}
\newcommand{\finalmtolbell}{$\log(M_*/L)=-0.55$}

\begin{document}

\title{Keck Observations of the Young Metal-Poor Host Galaxy of 
the Super-Chandrasekhar-Mass Type Ia Supernova SN~2007if}

\author{M.~Childress\altaffilmark{1,2}}
\author{G.~Aldering,\altaffilmark{1}}
\author{C.~Aragon,\altaffilmark{1}}
\author{P.~Antilogus,\altaffilmark{3}}
\author{S.~Bailey,\altaffilmark{1}}
\author{C.~Baltay,\altaffilmark{4}}
\author{S.~Bongard,\altaffilmark{3}}
\author{C.~Buton,\altaffilmark{5}}
\author{A.~Canto,\altaffilmark{3}}
\author{N.~Chotard,\altaffilmark{6}}
\author{Y.~Copin,\altaffilmark{6}}
\author{H.~K.~Fakhouri,\altaffilmark{1,2}}
\author{E.~Gangler,\altaffilmark{6}}
\author{M.~Kerschhaggl,\altaffilmark{5}}
\author{M.~Kowalski,\altaffilmark{5}}
\author{E.~Y.~Hsiao,\altaffilmark{1}}
\author{S.~Loken,\altaffilmark{1}}
\author{P.~Nugent,\altaffilmark{7}}
\author{K.~Paech,\altaffilmark{5}}
\author{R.~Pain,\altaffilmark{3}}
\author{E.~Pecontal,\altaffilmark{8}}
\author{R.~Pereira,\altaffilmark{6}}
\author{S.~Perlmutter,\altaffilmark{1,2}}
\author{D.~Rabinowitz,\altaffilmark{4}}
\author{K.~Runge,\altaffilmark{1}}
\author{R.~Scalzo,\altaffilmark{4,11}}
\author{R.~C.~Thomas,\altaffilmark{7}}
\author{G.~Smadja,\altaffilmark{6}}
\author{C.~Tao,\altaffilmark{9,10}}
\author{B.~A.~Weaver,\altaffilmark{12}}
\author{C.~Wu\altaffilmark{3}}

\altaffiltext{1}
{
    Physics Division, Lawrence Berkeley National Laboratory, 
    1 Cyclotron Road, Berkeley, CA, 94720
}
\altaffiltext{2}
{
    Department of Physics, University of California Berkeley,
    366 LeConte Hall MC 7300, Berkeley, CA, 94720-7300
}
\altaffiltext{3}
{
    Laboratoire de Physique Nucl\'eaire et des Hautes \'Energies,
    Universit\'e Pierre et Marie Curie Paris 6, Universit\'e Paris
    Diderot Paris 7, CNRS-IN2P3,  
    4 place Jussieu, 75252 Paris Cedex 05, France
}
\altaffiltext{4}
{
    Department of Physics, Yale University, 
    New Haven, CT, 06250-8121
}
\altaffiltext{5}
{
    Physikalisches Institut, Universit\"at Bonn,
    Nu\ss allee 12, 53115 Bonn, Germany
}
\altaffiltext{6}
{
    Universit\'e de Lyon, 69622, France; Universit\'e de Lyon 1, France;
    CNRS/IN2P3, Institut de Physique Nucl\'eaire de Lyon, France
}
\altaffiltext{7}
{
    Computational Cosmology Center, Computational Research Division,
    Lawrence Berkeley National Laboratory,  
    1 Cyclotron Road MS 50B-4206, Berkeley, CA, 94611
}
\altaffiltext{8}
{
    Centre de Recherche Astronomique de Lyon, Universit\'e Lyon 1,
    9 Avenue Charles Andr\'e, 69561 Saint Genis Laval Cedex, France
}
\altaffiltext{9}
{
    Centre de Physique des Particules de Marseille , 163, avenue de
    Luminy - Case 902 - 13288 Marseille Cedex 09, France 
}
\altaffiltext{10}
{
    Tsinghua Center for Astrophysics, Tsinghua University, Beijing
    100084, China
}
\altaffiltext{11}
{   Skymapper Fellow, current address:
    Research School of Astronomy \& Astrophysics, 
    Mount Stromlo Observatory,
    The Australian National University, 
    Cotter Road,
    Weston ACT 2611 Australia
}
\altaffiltext{12}
{
    New York University, Center for Cosmology \& Particle Physics,
    4 Washington Place, New York, NY, 10003
}

\begin{abstract}
We present Keck LRIS spectroscopy and $g$-band photometry of the
metal-poor, low-luminosity host galaxy of the super-Chandrasekhar mass
Type Ia supernova SN~2007if.  Deep imaging of the host reveals its
apparent magnitude to be \finalappmag, which at the
spectroscopically-measured redshift of \finalzhelio\
corresponds to an absolute magnitude of \finalabsmag. Galaxy $g-r$
color constrains the mass-to-light ratio, giving a host stellar mass
estimate of \finalhostmass.  Balmer absorption in the stellar
continuum, along with the strength of the 4000\AA\ break, constrain
the  age of the dominant starburst in the galaxy to be \finalhostage,
corresponding to a main-sequence turn-off mass of \finalhostmsto.
Using the R$_{23}$ method of calculating metallicity from the fluxes
of strong emission lines, we determine the host oxygen abundance to be
\finalmetalkk, significantly lower than any previously reported
spectroscopically-measured Type Ia supernova host galaxy
metallicity. Our data show that SN~2007if is very likely to have
originated from a young, metal-poor progenitor.
\end{abstract}


\section{Introduction}
Type Ia Supernovae (\sneia) are vital cosmological tools for measuring
the expansion history of the Universe \citep{42sne, riess98}, as their
luminosities show low intrinsic dispersion ($\sim 0.35$~mag), making
them ideal as distance indicators.  Outliers from the typical
luminosity distribution present an opportunity to explore the
underlying physical mechanism in these systems, and provide a
critical cross-check for possible ``contamination'' of future
high-redshift \snia\ surveys focusing on the normal \sneia.
Recently a potential new subclass of exceptionally overluminous
\sneia\ has been discovered, starting with the prototype SN2003fg
\citep[SNLS-03D3bb][]{howell06}, followed by 
SN~2007if \citep{scalzo10,yuan10} and SN2009dc \citep{tanaka10,
yamanaka09, silverman10, tauben10}, and potentially SN2006gz
\citep{hicken07}.

The generally accepted scenario for the production of a \snia\ is the
total disruption of a carbon-oxygen (CO) white dwarf (WD) by
thermonuclear runaway as accretion from a binary companion drives it
toward the Chandrasekhar mass ($M_{Ch}$).  In the
single-degenerate scenario \citep[SD;][]{whelan73}, the WD accretes
material from a less-evolved companion, either a main-sequence or
red giant star \citep[e.g.][]{hkn08}. In the double-degenerate
scenario \citep[DD;][]{iben84}, two WDs coalesce following orbital
decay from gravitational radiation. \citet{howell06} were the first to
suggest that this new subclass of overluminous \sneia\ are likely the
product of \superc-mass (SC) progenitors systems where substantially
more material than $M_{Ch}$ undergoes thermonuclear runaway, producing
more $^{56}$Ni \citep[see, e.g.,][]{raskin10} and resulting in a much
more luminous explosion. This interpretation is difficult to reconcile
with the traditional \snia\ progenitor scenarios in which the SN
itself is triggered as the WD approaches $M_{Ch}$.  In the SD scenario
where accretion onto the WD is posited to be steady and stable, an
accumulation of significantly more mass than $M_{Ch}$ is highly
unlikely \citep{piro08}.  In the DD scenario, the merger of two WDs
whose total mass exceeds $M_{Ch}$ (even by a significant amount) is a
natural  occurrence, and has made this scenario a favored framework
for interpreting the origin of \superc\ \sneia.  There are concerns,
however, that the merger of two WDs could result in accretion-induced
collapse rather than thermonuclear runaway
\citep[e.g.][]{nomoto95}. Independent constraints on the probable
progenitor properties of SC \sneia\ are therefore critical for
unraveling the mystery surrounding these exceptional SNe.

SN~2007if is particularly interesting among this new subclass of
probable \superc\ \sneia, as it has been shown to be the most luminous
\snia\ ever discovered, with a peak $V$-band magnitude of
$M_{V,07if}=-20.4$ \citep{scalzo10} -- nearly a full magnitude
brightner than the average \snia\ luminosity of $M_{V,Ia} \sim -19.5$
\citep{leibundgut00}. SN~2007if is also interesting for its extremely
faint host galaxy ($M_g \sim -14.5$), which we will show below is the
lowest-measured metallicity \snia\ host galaxy known.
SN~2007if was discovered by the ROTSE-III supernova search
\citep{akerlof07} on 2007 August 16.3 UT, and
independently by the Nearby Supernova Factory
\citep[SNfactory,][]{ald02} as SNF20070825-001 on 2007 August 25.4 UT
\citep[see][for details]{scalzo10}.  Located at
$\alpha_{2000}=$~01:10:51.37, $\delta_{2000}=$~+15:27:39.9, SN~2007if
showed no apparent host in search reference images, or in images from
the Sloan Digital Sky Survey \citep[SDSS;][]{york00}.  Our deep
co-add of NEAT + Palomar-QUEST search data showed a
potential host at $m_i \approx 23.3 \pm 0.4$ \citep{nugent07}, which
at the estimated redshift of SN~2007if would make its host galaxy
(hereafter HOST07if) one of the faintest \snia\ hosts ever discovered,
suggesting very low metallicity.

In this paper we present observations of HOST07if that confirm its
exceptionally low metallicity and establish its
luminosity-weighted age.  In \S\ref{sec:observations} we
present our photometry and spectroscopy, then describe the
derivation of gas-phase metallicity and stellar mass and age for
HOST07if in \S\ref{sec:host_metal} and \S\ref{sec:host_age},
respectively.  In \S\ref{sec:systematics} we describe cross-checks
performed to inspect potential biases or systematic effects in our
analysis. We discuss our results in the context of \snia\ hosts
and their implications for \snia\ progenitor scenarios in
\S\ref{sec:discussion}. \S\ref{sec:conclusions} presents our
conclusions. Throughout this paper we employ a standard $\Lambda$CDM
cosmology of $H_0=70$~km/s~Mpc$^{-1}$, $\Omega_m=0.3$, and
$\Omega_\Lambda=0.7$.

\section{Observations}
\label{sec:observations}

\begin{figure}[t]
\begin{center}
\includegraphics[width=0.45\textwidth]{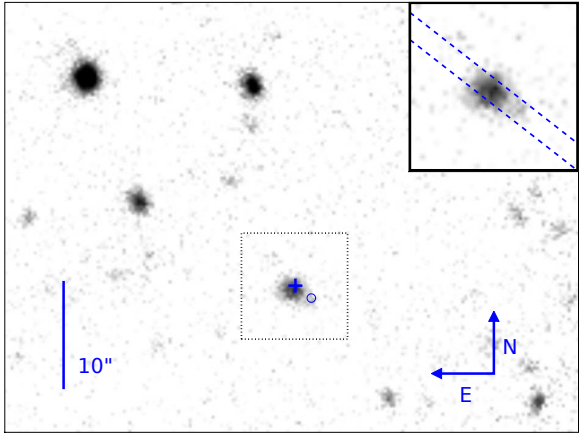}
\end{center}
\caption{Keck LRIS image of HOST07if.  The blue cross denotes the
  location of the supernova.  For reference, the ``bright'' field
  star in the upper left has magnitude $m_g = 21.1$.  The area
  immediately around HOST07if, denoted by the dotted box, is shown in
  the upper right inset along with the slit location shown as the
  dashed lines.  The high-redshift background galaxy appears just to
  the southwest of HOST07if, and its location is marked by the thin
  blue circle.}
\label{fig:host_image}
\end{figure}

HOST07if was observed with the Low Resolution Imaging Spectrometer
\citep[LRIS,][]{oke95} on the Keck I 10-m telescope on Mauna Kea on
2009 August 23 and 24 UT.  We employed the Keck-I atmospheric
dispersion corrector \citep[ADC;][]{keckadc}. On 2009 August 
23.6 five exposures of 100~s duration were obtained in imaging mode
using the blue camera of LRIS equipped with a $g$-band filter.  The
images were dithered to allow rejection of cosmetic defects, cosmic
rays, and to provide image coverage across the detector gap.  These
images were combined to form a deep image of HOST07if and
assess the potential for spectroscopic observation.  On the following
night (2009 August 24.6 UT) five additional imaging exposures of 100~s
duration were obtained in $g$-band to provide additional photometric
depth, then the target was aligned on the slit in imaging mode and the
instrument configured for spectroscopic observations. The blue side
was configured with the 600~l/mm grism blazed at 4000~\AA,
covering 3500-5600~\AA, and on the red side the 900~l/mm
grating blazed at 5500~\AA\ was employed, covering 5500-7650~\AA.  The
D560 dichroic beamsplitter was used, and no order-blocking filters
were necessary. A 1\arcsec\ slit was oriented at a position angle of
128$^{\rm o}$ along the apparent major axis of HOST07if, which
fortuitously was only a few degrees away from the parallactic
angle.  Our final co-added LRIS image for HOST07if is shown in
Figure~\ref{fig:host_image}, along with an overlay of the slit.  
Analysis of the acquisition and slit images show HOST07if to be
aligned on the center of the slit to within 1 pixel ($0\farcs27$). The
chosen slit gave resolutions of $\lambda/\Delta\lambda \sim 1000$
(4.4~\AA) and $\sim 1600$ (4.1~\AA) for the blue and red sides,
respectively.  Four spectroscopic exposures of 900~s duration were
obtained, starting at airmass 1.00 and ending at airmass 1.02. The
Keck-I ADC was employed, so we expect no chromatic slit loss due to
atmospheric differential refraction. Processing of the photometry and
spectroscopy are described below.

\subsection{Spectroscopy}
\begin{figure}[t]
\begin{center}
\includegraphics[width=0.48\textwidth]{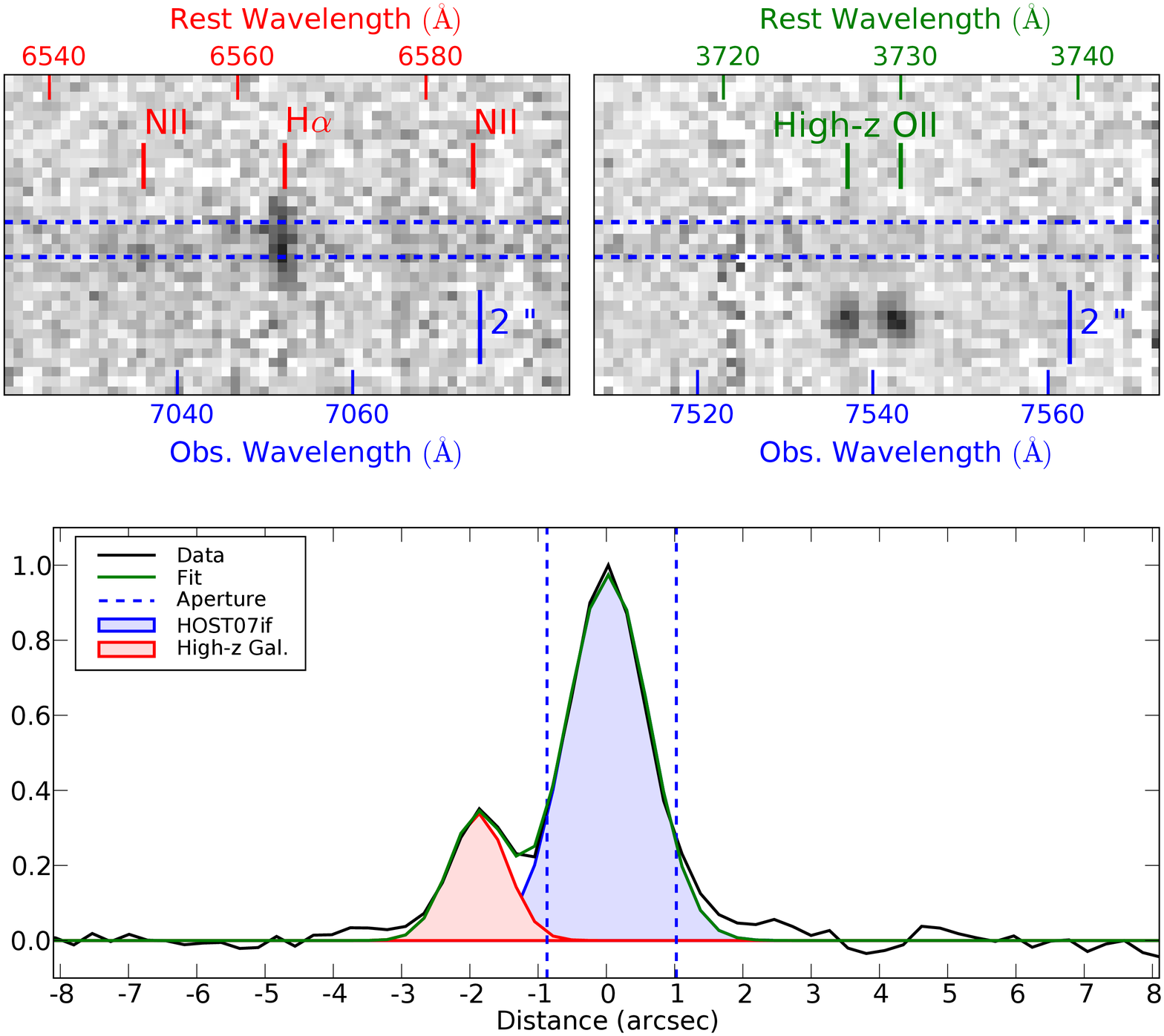}
\end{center}
\caption{\emph{Top:} Portions of the 2D sky-subtracted spectrum image
  showing (left) the strong \ha\ feature of HOST07if at
  $\lambda=7051$\AA\ corresponding to $z=0.074$ (note the distinct
  absence of \nii), and (right) the \oii\ feature of the high-$z$
  background galaxy at $\lambda\lambda7357,7543$\AA\ corresponding to
  $z=1.02$.  \emph{Bottom:} Wavelength-collapsed object profiles in
  $g$-band, showing our two-Gaussian fit to HOST07if and the high-$z$
  background galaxy, and the extraction aperture chosen for the
  HOST07if spectrum. Note the possible contamination from the high-$z$
  galaxy is extremely small.}
\label{fig:apertures}
\end{figure}

The LRIS spectra were reduced in IRAF\footnote{
IRAF is distributed by the National Optical Astronomy Observatory
which is operated by the Association of Universities for Research in
Astronomy, Inc., under cooperative agreement with the National Science
Foundation.}
using standard techniques.  Overscan subtraction was performed for
each of the four amps, and the data were mosaiced to form individual
two-dimensional frames with data from each amp scaled by its gain.
We subtracted bias frames from these data, removed cosmic rays using
{\tt LA Cosmic} \citep{vandokkum01}, and removed pixel variations in
detector efficiency by dividing images by wavelength-normalized flat
field dome lamp exposures. The two-dimensional wavelength solution for
the blue channel was derived from nightly arc lamp exposures with a
linear shift in wavelength applied by measuring the [OI]~$\lambda5579$
atomic night sky line. This linear shift was verified by
cross-correlation of the sky spectrum with a high-resolution night sky
spectrum from \citet{uves}. For the red channel, two-dimensional
wavelength solutions for object exposures were derived from night sky
lines in the object exposures, while for standard stars we used
nightly arc lamp exposures with a wavelength shift determined from
[OI]~$\lambda\lambda6300,6364$ sky lines. Object spectra were reduced
to one dimension using the IRAF function {\tt apall}, and nightly flux
calibrations were derived from standard stars observed at appropriate
ranges of airmass. Telluric absorption features were then removed
using the nightly standard star spectra.  Finally, the spectrum was
corrected for observer motion with respect to the heliocentric frame,
and the Galactic reddening of the spectrum was corrected using the
\citet{ccm} law and the value $E(B-V) = 0.079$ \citep{sfd98}.

The two-dimensional spectrum of HOST07if showed the presence of a
background galaxy separated from HOST07if by $1\farcs9$ and
displaying a strong \oii\ doublet at $\lambda\lambda7537,7543$~\AA,
corresponding to $z=1.02$.  Correction for this object in photometric
measurements will be described below.  We show portions of the
background-subtracted 2D red side spectroscopy image in the top panel
of Figure~\ref{fig:apertures} to show the offending emission lines
from the high-$z$ object.  The lower panel of the same figure shows
the wavelength-collapsed spatial profile of the 2D blue side
spectroscopy image along with the chosen extraction aperture.  Based
on profile fits to the two objects, we estimate the possible
contamination of the extracted HOST07if spectrum by
the high-$z$ object to be less than 0.5\% at all wavelengths (except
at the high-$z$ [OII] doublet position, which does not affect any
emission line measurements for HOST07if).

\subsection{Photometry}
\label{sec:host_phot}
LRIS blue channel photometry was processed in IRAF.  Overscan
subtraction and mosaicing were performed in the same manner as for the
spectroscopy, except that blank pixels were inserted between data from
the two detectors to account for the physical gap between the two
chips. The images were flat-fielded using $g$-band dome flats taken
earlier in the night.  Astrometric solutions were derived using {\tt
WCSTools} \citep{mink06}, then refined using {\tt SCAMP}
\citep{bertin06} matching to 2MASS \citep{twomass}. Individual
exposures were combined with {\tt SWARP} \citep{bertin02} using median
addition, and with proper de-weighting of the detector gap regions and
weighting of images by exposure time.  With the 5$\times$100~s
exposures of 2009-08-23 UT, 5$\times$100~s exposures of 2009-08-24 UT,
and the 4$\times$60~s exposures used for target alignment, the total
imaging time at the target location is 1240~s.

The photometric zeropoint for the target was derived by matching
objects in the field to SDSS \citep{york00} photometry.  We extracted
magnitudes for all objects in the field using {\tt SExtractor}
\citep{sextractor} using the {\tt MAG\_AUTO} output parameter, which
measures the flux inside an elliptical Kron-like aperture.  We then
matched objects in our field to the SDSS DR7 \citep{abaz09} {\tt
  PhotoObjAll} $g$-band model magnitudes, ensuring the photometry was
clean and the objects were primary targets ({\tt mode=1}). Given the
depth of the LRIS imaging, targets brighter than $m_g \sim 16.5$
saturated the detector, so we chose the bright magnitude limit of our
catalog matching to be $m_g \sim 17.0$. The SDSS $g$-band completeness
limit is estimated at $m_g \sim 22.2$ with deviation from Poggson
magnitudes beginning at about $m_g \sim 22.6$ \citep{stoughton02}, so
we conservatively chose a magnitude limit of $m_g \sim 22.0$ for our
catalog matching.  We therefore calculate the photometric zeropoint
using the error-weighted mean of $N=20$  objects between $17.0 < m_g <
22.0$ and find $m_{ZP} = 32.59 \pm 0.04$.  

The raw instrumental $g$-band magnitude for HOST07if was observed to
be $m_{g,inst}=-9.38\pm0.03$. Combined with the SDSS zeropoint and
error, we determined the raw $g$-band magnitude of HOST07if as
observed with LRIS to be $m_g = 23.21 \pm 0.05$.  The Galactic
reddening of $E(B-V) = 0.079$ \citep{sfd98} results in a $g$-band
extinction of $A_g = 0.34$.  In our stacked 
image, HOST07if is blended with the background high-redshift galaxy
described above.  To account for its contribution to the measured
HOST07if flux, we analyze the two-dimensional blue channel spectrum,
which was taken during the best seeing conditions of both nights
($\sim$ 0\farcs6) and shows a clear separation of the two objects.
We subtract the sky background from the 2D spectrum, apply the flux
calibration and multiply by the $g$-band filter throughput in the
wavelength direction, then collapse the 2D spectrum in wavelength
along the aperture trace.  This effectively provides a high
signal-to-noise measurement of the object profiles along the slit
direction in $g$-band.  We then fit this 1D profile with two
Gaussians; the data and fit are shown in Figure~\ref{fig:apertures}
along with the chosen aperture.  The center of both objects fall
inside the slit and
the seeing was smaller than the slit width, so we predict that the
flux within the slit satisfactorily preserves the flux ratio between the
two objects. The ratio of the flux of the high-$z$ galaxy to
HOST07if in $g$-band is $F_\mathrm{high-z}/F_\mathrm{HOST07if} = 0.27
\pm 0.02$,
with a separation of $1\farcs9$. This results in a correction to the
observed magnitude of HOST07if of $\Delta m_g = -0.26 \pm 0.02$.
Finally we include the known offset between the SDSS and AB magnitude
systems \citep{stoughton02} of $m_{g,AB} = m_{g,SDSS} + 0.02$.  To
derive the rest-frame $g$-band magnitude, we perform a K-correction
\citep{nkp02} using the $g$-band filter throughput and the HOST07if
spectrum, finding $K_g = -0.002$. The reddening, object overlap,
SDSS-AB offset, and K-correction effects result in a final rest-frame
$g$-band magnitude of HOST07if of \finalappmag.

To derive the correct distance modulus for HOST07if, we convert the
heliocentric redshift derived from nebular emission lines (see
\S\ref{sec:emline_fluxes}) to the CMB rest frame using the 
dipole parameters from WMAP5 \citep{hinshaw09} to obtain
\finalzcmb. Assuming standard $\Lambda$CDM cosmology
($H_0=70$~km/s~Mpc$^{-1}$, $\Omega_m=0.3$, $\Omega_\Lambda=0.7$), we
use the code of \citet{wright06} to calculate a distance modulus of
\finaldistmod\ \citep[note this corrects a transcription error in the
calculation of the host absolute magnitude reported in][ which did
not affect any other values reported in that analysis]{scalzo10}. With
the apparent magnitude  derived above, this gives HOST07if an absolute
$g$-band magnitude of \finalabsmag. 

Since the LRIS $g$-band observations were the only deep late-time
photometry of the host (after the SN had fully faded), we analyze the
HOST07if spectrum as a source of galaxy color information.  We
synthesize rest-frame $u$- $g$- and $r$-band magnitudes from the
spectrum using the SDSS filter transmissions\footnote{The 
SDSS filter transmissions are available at 
{\tt http://www.sdss.org/dr7/instruments/imager/index.html}.}
and obtain effective observer-frame galaxy colors of \finalgrcolor\
and \finalugcolor. The relative flux calibration of our spectrum is
very good, as we measure the synthetic $g-r$ and $u-g$ colors of the
night's standard star observations to match those synthesized from
calibration spectra to within $\Delta(g-r) < 0.01$~mag and
$\Delta(u-g) < 0.01$~mag, primarily driven by noise in the dichroic
region. These colors will be used below to derive the galaxy
mass-to-light ratio and to inspect possible reddening due to dust.

\section{SN~2007if Host Metallicity}
\label{sec:host_metal}
Our original objective in observing HOST07if was to secure a host
redshift in order to accurately determine the SN~2007if ejecta
velocity. This measurement played a key role in establishing the
kinetic energy of the explosion and SN~2007if as having a mass greater
than the Chandrasekhar limit \citet{scalzo10}. Fortuitously, the final
spectrum showed emission in \ha\ and \oii\ sufficiently strong to
measure a gas-phase metallicity.

\subsection{Emission Line Fluxes}
\label{sec:emline_fluxes}
\begin{figure*}[t]
\begin{center}
\includegraphics[width=0.90\textwidth]{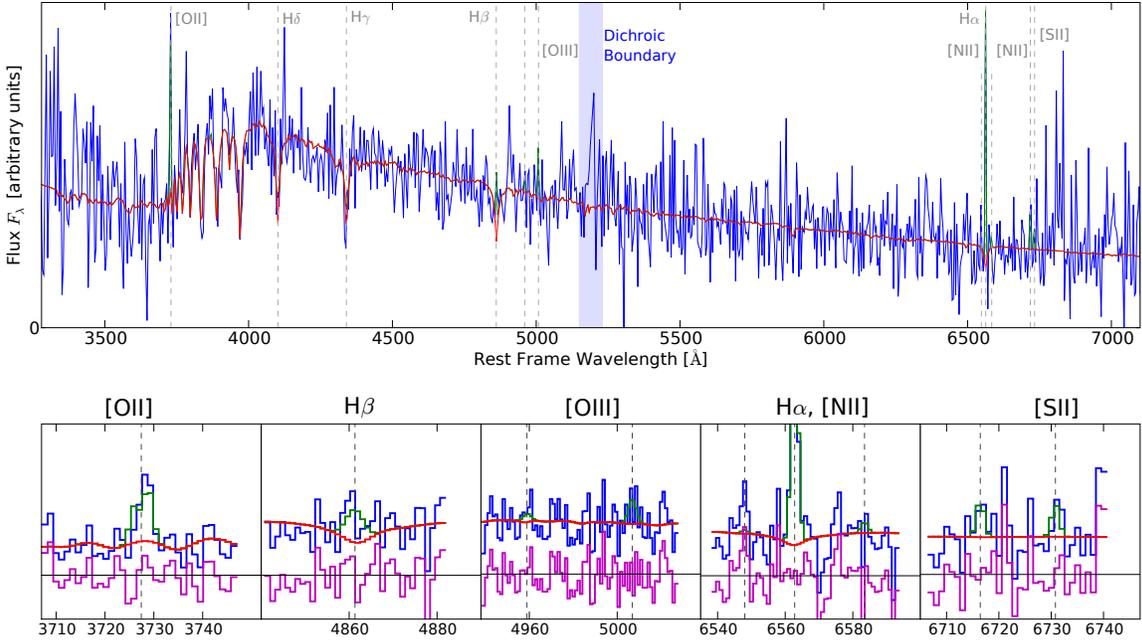}
\end{center}
\caption{\emph{Top:} Spectrum of HOST07if (blue) binned to 4\AA\ for
  visual clarity, with fitted background (red) and emission lines
  (green). \emph{Bottom:} Zoomed fit regions for notable emission
  lines (unbinned), with fit residuals (magenta).}
\label{fig:spectrum}
\end{figure*}

Accurate measurement of emission line fluxes in star-forming galaxies
requires proper accounting for stellar absorption.  To this end we fit
the emission line fluxes and stellar background in the HOST07if
spectrum simultaneously using a modified version of the IDL routine
{\tt linebackfit} from the 
{\tt
idlspec2d}\footnote{http://spectro.princeton.edu/idlspec2d\_install.html} 
package developed by the SDSS team.  
This routine allows the user to provide a list of template spectra fit in
linear combination with Gaussian emission line profiles.  We have modified
this code to force the background coefficients to be non-negative and have
incorporated the ability to fit for a scaling factor between the blue
and red channels of two-arm spectrograph data.  For background
templates we chose a set of simple stellar populations (SSPs) from the
stellar population synthesis code GALAXEV \citep[][BC03]{bc03} with a
\citet{chab03} IMF and the same time sampling used for background
fitting by \citet[][T04]{trem04}, which ultimately consists of ten
SSPs for each metallicity. We note that the use of \citet{salpeter}
IMF templates results in negligible differences to the fitted emission
line fluxes, and metallicity difference smaller than the quoted
precision of 0.01~dex.

We fit the two LRIS channels simultaneously, with the background
templates on each channel convolved to the spectrograph resolution for
each channel, namely 4.4~\AA\ and 4.1~\AA\ for the blue and red channels
respectively, and fit the cross-channel scaling simultaneously. As
with the SDSS spectroscopic pipeline, our emission line fitting is
done in an iterative fashion.  An initial guess of the redshift is
used to set the redshift of the background templates, and the spectrum
is fit with the widths and redshifts of all lines allowed to float
freely. The best redshift is measured from the initial emission line
fits, and a second iteration is performed with the redshift of the
background templates set to this value. The emission line fluxes are
then measured with the line redshifts all fixed to this value. We found
the best fit $\chi^2$ was obtained when the background templates were
drawn from the $Z=0.004$ track. The uncertainty in the scaling of the
blue and red channels is measured to be $\approx 3\%$, and has a value
consistent with those measured for our standard stars. The
uncertainties from all fit parameters and their covariances are
measured by the fitting code, and emission line flux errors accurately
reflect the influence of all fit parameters in their estimation
(including the cross-channel scaling).  The final emission line fluxes
from our best fit are presented in Table~\ref{tab:emlines}, and the
fit to the spectrum is shown in Figure~\ref{fig:spectrum}.

\begin{table}[h]
\centering
\caption{HOST07if Emission Line Fluxes}
\centering
\begin{tabular}{ l r r }
\hline
Line     & Obs. Flux\tablenotemark{1} & 
$F(\lambda)/F(H\beta*)$\tablenotemark{2} \\
\hline
\oii\       & $48.80 \pm 10.56$ & $2.44 \pm 0.53$ \\
\hb\        & $22.29 \pm  7.17$ & $1.11 \pm 0.36$ \\
\oiiia\     & $ 5.22 \pm  2.85$ & $0.26 \pm 0.14$ \\
\oiiib\     & $15.37 \pm  8.38$ & $0.77 \pm 0.42$ \\
\ha\        & $57.46 \pm  5.30$ & $2.87 \pm 0.26$ \\
\niia\      & $ 0.93 \pm  1.25$ & $0.05 \pm 0.06$ \\
\niib\      & $ 2.77 \pm  3.72$ & $0.14 \pm 0.19$ \\
\siia\      & $ 7.46 \pm  3.52$ & $0.37 \pm 0.18$ \\
\siib\      & $ 6.68 \pm  3.71$ & $0.33 \pm 0.19$ \\
\hline
\end{tabular}
\tablenotetext{1}{
Fluxes in units of $10^{-19} ergs\cdot cm^{-2}\cdot s^{-1}$} 
\tablenotetext{2}{
$F(H\beta*)\equiv F(H\alpha)/2.87$; see text for details.}
\label{tab:emlines}
\end{table}

The emission lines from our spectrum of HOST07if provide a formal
redshift and uncertainty of \finalzlines\ in the heliocentric frame.
This value is slightly different from the value we quoted in
\citet{scalzo10}, and reflects a more thorough treatment of the
spectrum wavelength solution. Additionally, we calculate the
contribution of our wavelength solution to the redshift error budget
to be $\Delta z_\mathrm{wsol} \approx 2.5\times10^{-5}$. Because our
object has extent smaller than the slit, the dominant source of
redshift error from our data comes from the centering of the object on
the slit. As stated above, we measure this error to be no more than 1
pixel, which corresponds to a redshift error of $\Delta
z_\mathrm{slit} \approx 1.5\times10^{-4}$ at \ha, the line which best
constrains the redshift. Thus we estimate the final heliocentric
redshift and error for HOST07if to be \finalzhelio.

The Balmer emission line fluxes are typically used to estimate
intrinsic reddening in galaxies by comparison to the Case B
recombination value of $F(H\alpha)/F(H\beta) = 2.87$ at a temperature
of $T=10,000K$ \citep{agn2}. This value is well within the $1\sigma$
estimate from our measured emission line fluxes (0.62 in the
cumulative probability function), but is poorly constrained due to the
relatively low S/N of our spectrum. We therefore will report results
derived under the assumption of \emph{no intrinsic extinction}.
Later, in \S\ref{sec:reddening}, we show that this assumption is
supported by multiple facets of the data themselves, and even in the
worst case scenario of leaving reddening unconstrained has negligible
impact on our final results.

\subsection{Gas-Phase Metallicity}
Initial signs that HOST07if is a low metallicity galaxy include the
non-detection of \nii\ (below the noise threshold, see value and
errorbar in Table~\ref{tab:emlines} and 2D spectroscopic image in
Figure~\ref{fig:apertures}), the relatively weak \oii\ and \oiii\
lines (compared to the strong Balmer lines), and of course its
low luminosity. Low-metallicity galaxy abundances are
ideally determined using the ``direct'' method whereby the ratio of
the auroral \aoiii\ line flux to that of the stronger \oiii\ lines is
used to constrain the electron temperature $T_e$ in the doubly-ionized
oxygen (O$^{++}$) zone ($T_e$(OIII)). Because the auroral line is not
detected in HOST07if, and the intrinsically stronger \oiii\ lines are
only weakly detected, the direct method is untenable here.  

The question of appropriate metallicity scales will be addressed
later in \S\ref{sec:metal_scales}, but here we derive the metallicity
using the $R_{23}$ method of \citet[][hereafter KK04]{kk04}. The ratio
$R_{23}$ is double valued with metallicity, and the flux ratio
[NII]/H$\alpha$ is typically used to break the degeneracy and select
which ``branch'' of the $R_{23}$ metallicity calibration is
appropriate. For HOST07if, [NII]/H$\alpha$ indicates the lower
metallicity branch, so we employ the lower branch of the KK04
calibration of the $R_{23}$ method as updated by \citet{ke08}. This
method is advantageous because it iteratively calculates the
metallicity and ionization parameter.

To derive a tighter constraint on the metallicity of HOST07if, we use
the higher S/N \ha\ flux measurement and its error scaled by the
fiducial Balmer decrement as proxies for the flux and error of \hb. As
stated above, this is consistent with our assumption of no reddening
in HOST07if and results in an \hb\ flux only $0.25\sigma$ different
from that measured, but with an error bar $4\times$ smaller.
For HOST07if, we measure a metallicity of \finalmetalkk, with an
ionization parameter \finalionpar. This low value of the ionization
parameter is unsurprising given the strength of \oii\ and the relative
weakness of \oiii. These indicate that the ionizing radiation is dilute
and it has been some time since HOST07if's most recent burst of
star-formation (consistent with stellar absorption strengths -- see
below).  We note that \nii\ is used to break the $R_{23}$ degeneracy,
and our measurement of this line predicts the lower branch at only
$\approx69\%$ probability, since [NII] appears to be below the noise
level.  If we were to choose the upper $R_{23}$ branch, this would
make HOST07if a $>5\sigma$ outlier on the mass-metallicity relation
\citep[T04,][]{ke08}, an extremely rare event \citep[see
e.g.][]{peeples08}.  Additionally, [NII]/[OII] at such a high
metallicity \citep{kd02} would predict an [NII]$\lambda6584$ flux
strong enough to be detected at $>8\sigma$.

\section{SN~2007if Host Age and Stellar Mass}
\label{sec:host_age}
Information about the star-formation history (SFH) of HOST07if is
desirable for constraining the age of the SN~2007if progenitor.
Spectral indices measured from galaxy stellar spectra can be useful in
assessing the mean stellar age, likelihood of recent starburst, and
stellar mass-to-light ratios \citep[see e.g.][]{kauff03a, ez_ages,
gb09}. To facilitate the inspection of the SFH of HOST07if, we measure
several age-sensitive spectral indices from the emission-subtracted
spectrum of HOST07if and compared their values to model spectra
generated using stellar population synthesis (SPS) techniques. The
details of our analysis are as follows.

We measured the strength of several Balmer absorption features
according to their standard definition on the Lick system
\citep{worthey94,worthott97} as well as the strength of the 4000~\AA\
break \citep[D4000,][]{D4000} in the spectrum of HOST07if after
removing emission line features as determined in
\S\ref{sec:emline_fluxes}. Using the formulae of \citet{cardiel98},
we measure the values and errors reported in 
Table~\ref{tab:spectral_indices}. These indices are known to have
strong dependence on stellar age \citep{vazdekis10}, with negligible
dependence on instrument resolution (and by extension galaxy velocity
dispersion). The low D4000 and strong Balmer absorption we measure for
HOST07if is indicative of young stellar ages of a few hundred Myr.

\begin{table}[h]
\centering
\caption{HOST07if Spectral Indices}
\centering
\begin{tabular}{ l r }
\hline
Index       & Value \\
\hline
D4000       & $1.13 \pm 0.05$   \\
H$\delta_A$ & $3.50 \pm 2.33$   \\
H$\gamma_A$ & $7.19 \pm 2.36$   \\
H$\beta$    & $2.34 \pm 2.82$   \\
\hline
\end{tabular}
\label{tab:spectral_indices}
\end{table}

\begin{figure}[t]
\begin{center}
\includegraphics[width=0.45\textwidth]{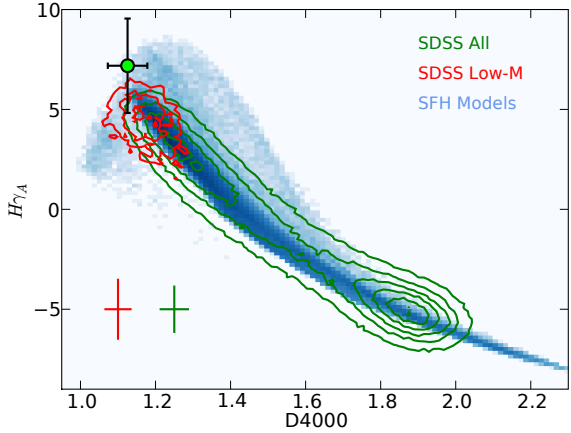}
\end{center}
\caption{Location of HOST07if (green circle) in the $H\gamma_A$-D4000
  plane compared to the library of physically motivated SFHs of GB09
  (blue background). Overplotted are index values for SDSS galaxies
  derived by the MPA-JHU group for the full galaxy mass range (green
  contours) as well as low mass ($\log(M_*/M_\odot) \leq 9.0$)
  galaxies (red contours), with median measurement errors shown as the
  colored crosses in the lower left.  Galaxies in the
  densely-populated band spanning the full range of D4000 have SFHs
  dominated by continuous star-formation, while the galaxies located
  away from this band have undergone recent burst of star-formation.}
\label{fig:sfh_diagram}
\end{figure}

To assess the general behavior of the SFH of HOST07if, we generate a
library of synthetic galaxy spectra using the BC03 SPS code and a
suite of physically-motivated SFHs. We follow the same prescription as
\citet{gallazzi05} and \citet[][hereafter GB09]{gb09} to generate
models consisting of an exponentially declining continuous SF
component superposed with random burst of SF (see GB09 for details).
We measured the same spectral indices from our model spectra and plot
the location of HOST07if and our model galaxies (blue background) in
H$\gamma_A$-$D4000$ space in Figure~\ref{fig:sfh_diagram}. For
reference, we also plot the location of SDSS DR7 galaxies whose
spectral index values and stellar masses have been measured by the
MPA-JHU group\footnote{http://www.mpa-garching.mpg.de/SDSS/DR7/}. The
full sample of galaxies between redshifts $0.005 \leq z \leq 0.25$ are
shown as the green contour, while low mass ($\log(M_*/M_\odot) \leq
9.0$) galaxies are shown as the red contour, with the median error
bars for each quantity (for each subsample) shown as the crosses in
the lower left.

\citet{kauff03a} showed the Balmer-D4000 diagram to be an informative
parameter space in which to inspect the SFH of star-forming galaxies,
and (GB09) extensively analyzed the properties of galaxies in
different regions of this diagram (for H$\delta_A$). The dense band of
model spectra (dark blue) and the majority of the SDSS galaxies form a
sequence of galaxies dominated by continuous star-formation ranging
from very old (high D4000, low H$\gamma_A$) to very young (low D4000,
high H$\gamma_A$) mean stellar ages. Galaxies whose indices are
located away from this band have undergone a strong starburst in the
past few hundred Myr.

It is evident that HOST07if is located away from the continuous SFH
band in this spectral index parameter space, and is even separated
from the majority of low mass galaxies whose mean stellar ages are
very young. This indicates that HOST07if underwent a major burst of
star formation in its recent past. This is perhaps unsurprising given
that HOST07if's low luminosity implies a low stellar mass, and low
mass dwarf galaxies tend to have SFHs characterized by strong yet
intermittent bursts of star-formation \citep{ss72}. In the case of 
a strong recent starburst, the light from the burst tends to dominate
the galaxy spectral energy distribution (SED), which can make it more
difficult to constrain the complete galaxy SFH and mass-to-light ratio
(GB09). Thus we will proceed by decoupling the recent burst of SF from
the remaining SFH of HOST07if. We will first assess the age of the most
recent starburst, then investigate the potential presence of older
stars in HOST07if.

\subsection{HOST07if Burst Stellar Age}
\label{sec:sfh}
To quantify the age of the most recent starburst in HOST07if, we
compare the HOST07if spectral indices to those of a library of
starburst model spectra generated from the BC03 SPS models. The burst
SFHs are simple boxcar functions in time described only by the start
and end time of the burst of star-formation. Burst start times are
uniformly distributed between 0 and 13.5~Gyr ago, and durations are
uniformly distributed between 10~Myr and 1~Gyr. Metallicities were
distributed logarithmically between $0.2 < Z/Z_\odot < 2.5$ and
distributed as a smoothly decaying function in metallicity ($\propto
\log(Z)^{1/3}$) between $0.02 < Z/Z_\odot < 0.2$ (in order to not
over-represent low-metallicity bursts).

We derive the luminosity-weighted HOST07if starburst age probability
distribution function (PDF) in a probabilistic fashion. For each
template galaxy in the burst library, we computed the values of the
spectral indices measured in the same way as HOST07if.  We then derive
each template's error-normalized separation from HOST07if in this
multi-dimensional parameter space defined by the spectral indices as
\begin{equation}
\chi^2_i = \sum_\alpha \left[\frac{a_{\alpha,i} - a_{\alpha,07if}}
{\sigma_{a_{\alpha,07if}}}\right]^2 
\end{equation}
where $a_{\alpha,i}$ is the value of parameter
$\alpha$ for template $i$, and similarly $a_{\alpha,07if}$ and
$\sigma_{a_{\alpha,07if}}$ 
are the value and uncertainty of that same parameter
for HOST07if.  Each template spectrum is a linear combination of 
spectra of SSPs of discrete age and metallicity as defined by the BC03
models.  We assign a weight to each SSP equal to its integrated
optical flux (3500~\AA\ $< \lambda <$ 10000~\AA), as the brighter SSPs
are more likely to drive the spectral features. Thus each template has
a luminosity-weighted age PDF that is the product of the template's
coefficients for each SSP multiplied by the luminosity weights for
each SSP (and normalized to unity probability). For each age bin in
the HOST07if burst age PDF, each template adds probability to the bin
that is a product of the template's age PDF value for that age bin and
the appropriate weighting ($\exp[-\chi_i^2/2]$) for the template's
parameter space separation from HOST07if.  The final burst age PDF for
HOST07if was renormalized according to the total probability of all
templates ($\sum_i\exp[-\chi_i^2/2]$). We use the
final HOST07if burst age PDF to derive the median age and $\pm1\sigma$
errors for the stellar population of HOST07if as derived from the
cumulative probability function.

We examined the accuracy of this method by performing the same
age measurement with our burst library SEDs where $g-r < 0.5$~mag
(thus the youngest subsample of bursts), which we will refer to as the
``validation sample''. We tested our method for a
variety of combinations of spectral indices, and measured the mean
offset from the true value (bias) and dispersion (systematic error)
for each combination.  In general, the bias was much smaller than the
dispersion, and the dispersion decreased as more Balmer indices were
added but saturated at the dispersion using the combination of
H$\delta$, H$\gamma$, and H$\beta$. The H$\beta$ absorption strength
for HOST07if is roughly the same magnitude as the emission equivalent
width ($EW(H\beta)=5.3\pm1.5$\AA), so the potential for emission
contamination of this index exceeds the reduction in systematic error
gained by its inclusion. Additionally, the age sensitivity of the
H$\beta$ index is slightly dependent on spectrograph resolution and
galaxy velocity dispersion \citep{vazdekis10}, whereas the other two
Balmer indices are not, and the velocity dispersion of HOST07if is
poorly constrained at our spectrum S/N. We thus exclude the H$\beta$
index from our parameter space. We also considered other (non-Balmer)
Lick indices, including G4300, but these provided no stronger
constraints on the HOST07if age.  Given the relative insensitivity of
these other indices in the $\sim$100~Myr age range found for HOST07if
\citep{vazdekis10}, this is unsurprising. 

\begin{figure}[t]
\begin{center}
\includegraphics[width=0.45\textwidth]{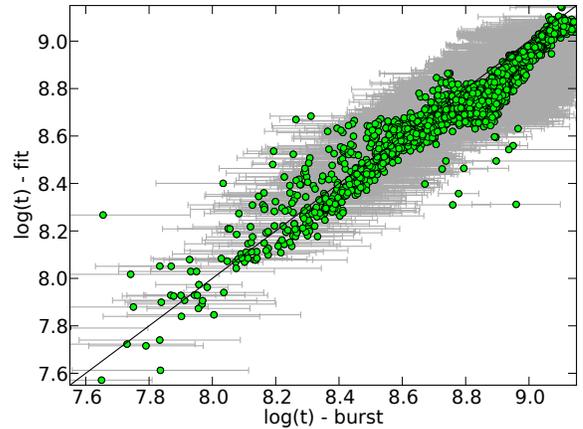}
\end{center}
\caption{Reconstructed starburst age for our selected library galaxies
vs. the input time of most recent starburst (green circles), with
duration of starburst shown as horizontal grey bars.  The scatter
about the true value is 0.06 dex.}
\label{fig:recon_validation}
\end{figure}

The final set of indices used to define the parameter space for
template matching was D4000, H$\delta$, and H$\gamma$. We show
in Figure~\ref{fig:recon_validation} the comparison between the median
reconstructed stellar age against the median time (green circles) and
duration (grey horizontal bars) of the starburst for each model in the
aforementioned validation sample. The final mean offset between input
and reconstructed age is $\Delta\log(t) = -0.05$~dex with a scatter of
$0.06$~dex.  We thus consider our reconstruction method to be
accurate, with a systematic age uncertainty of $\Delta\log(t) =
0.06$.

\begin{figure}[t]
\begin{center}
\includegraphics[width=0.45\textwidth]{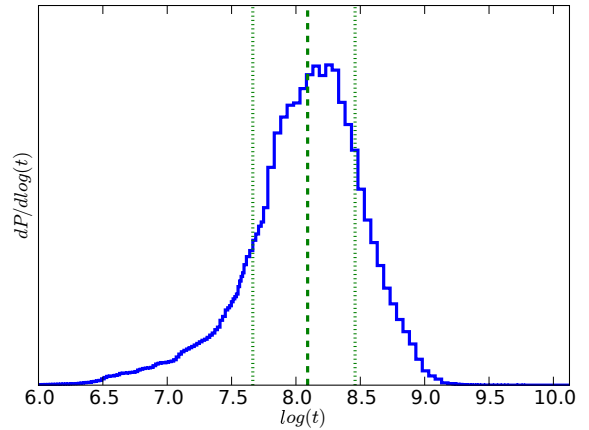}
\end{center}
\caption{Final luminosity-weighted burst age PDF for HOST07if.  The
  solid vertical line represents the median of the cumulative
  probability distribution, while the two dashed vertical lines
  represent the 16th and 84 percentile (i.e. $1\sigma$) of the same.}
\label{fig:age_pdf}
\end{figure}

The final burst age PDF for HOST07if is shown in Figure
\ref{fig:age_pdf}, and we can see that the age constraint is
remarkably tight.  Our analysis shows the luminosity-weighted stellar
age of HOST07if to be \finalhostlogage, or in linear age
\finalhostage\ (with the addition of statistical and systematic errors
in quadrature). For the BC03 tracks at metallicity $Z=0.004$ (the
closest value to our derived galaxy gas-phase metallicity) this
corresponds to a main-sequence turn-off mass of \finalhostmsto.

We also investigated the inclusion of optical colors in constraining
stellar age, beginning with $g-r$ which is strongly correlated with
stellar age and was shown by GB09 to be a good color for constraining
$M_*/L$. The age implied by the HOST07if $g-r$ optical color was
somewhat in tension with that implied by the spectral indices
($\log(t) \sim 8.5$ vs. $\log(t) \sim 8.1$). The cause of this
discrepancy is likely to be either the presence of some older stars in
addition to the recent burst of star-formation (see
\S\ref{sec:old_stars}), or intrinsic reddening in the galaxy (see
\S\ref{sec:reddening}).

It is worth noting that the distribution of ages and metallicities in
the model library has a significant impact on the resultant age
PDF. Applying the same method to the GB09-like SFH library used in
Figure~\ref{fig:sfh_diagram} yields a very differently shaped PDF, and
does not successfully recover burst ages for those SFHs with a
dominant recent starburst. This is because the manner in which the
library SFHs populate the age-metallicity parameter space effectively
acts as a prior on the resultant age PDF, whose final form is
especially dependent on the way in which different age bins are
coupled to one another by the assumed shape of the SFH. Our burst
model library employs the simplest possible SFHs (excepting of course
a $\delta$-function SFH) and provides an effectively flat and
decoupled prior because it populates age and metallicity bins 
evenly and only couples adjacent age bins with equal weight and over
relatively short ($<1$~Gyr) timescales.

Finally, we note that our burst age assessment method also provides 
corroboration of the low metallicity of HOST07if (measured from
emission lines above in \S\ref{sec:host_metal}). In addition to
tracking the age distribution of each template, we can inspect the
distribution of metallicity tracks used in construction of the
templates. Thus we can examine the burst age PDF as a function of
metallicity, and derive the integrated probability for each BC03
metallicity track. Doing so yields the following probabilities: 25\%
for $Z=0.0004$, 40\% for $Z=0.004$, 17\% for $Z=0.008$, 13\% for
$Z=0.02$ (solar), and 5\% for $Z=0.05$. This discrete distribution
illustrates the strong preference for lower metallicity tracks despite
the relatively flat prior (w.r.t. each track). This is a product of
the metallicity sensitivity of the spectral indices used in the
data-model comparison, and shows that the stellar spectral features
favor a low metallicity in agreement with our measurement of the gas
phase metallicity above.

\subsection{Old Stars in HOST07if}
\label{sec:old_stars}
Perhaps the greatest limitation in our ability to constrain the age of
the SN~2007if progenitor is the uncertainty in the amount of old stars
in HOST07if. Low-mass dwarf galaxies such as HOST07if are likely to
have a bursty SFH \citep{ss72} characterized by intense bursts of
star-formation separated by extended quiescent periods of reduced SFR
\citep{sa08}. Such galaxies may have formed the majority of their
stars in the distant past \citep{zgg11}, so it is critical to
investigate the potential amount of old stars in HOST07if. 

The bursty nature of the HOST07if SFH is supported by the comparison
of the burst star-formation rate (SFR) implied by our age constraint
as compared to that implied by the observed \ha\ emission. We showed
above that the HOST07if spectrum is dominated by stars of age
\finalageapprox, and we can make a simple approximation of the mass of
stars formed during the burst by multiplying the observed $g$-band
flux by the mass-to-light ratio of our estimated burst age (and
metallicity). Doing so yields an approximate mass of $10^7M_\odot$ of
stars formed in the burst, and if we assume this was formed in
$t\approx100$~Myr (likely an extreme over-estimate), we can estimate a
rough burst SFR of \finalsfrburst. The presence of \ha\ emission
implies some current star-formation, which we can quantify using the
formula of \citet{kenn98} to find \finalsfrhalpha. Thus, even our
crude estimate of the burst SFR shows the ratio of SFR during the
burst to that at the present time to be at least \finalsfrratio, which
implies that the HOST07if SFR is tapering off from its intense value
during the recent burst.

\begin{figure}[t]
\begin{center}
\includegraphics[width=0.45\textwidth]{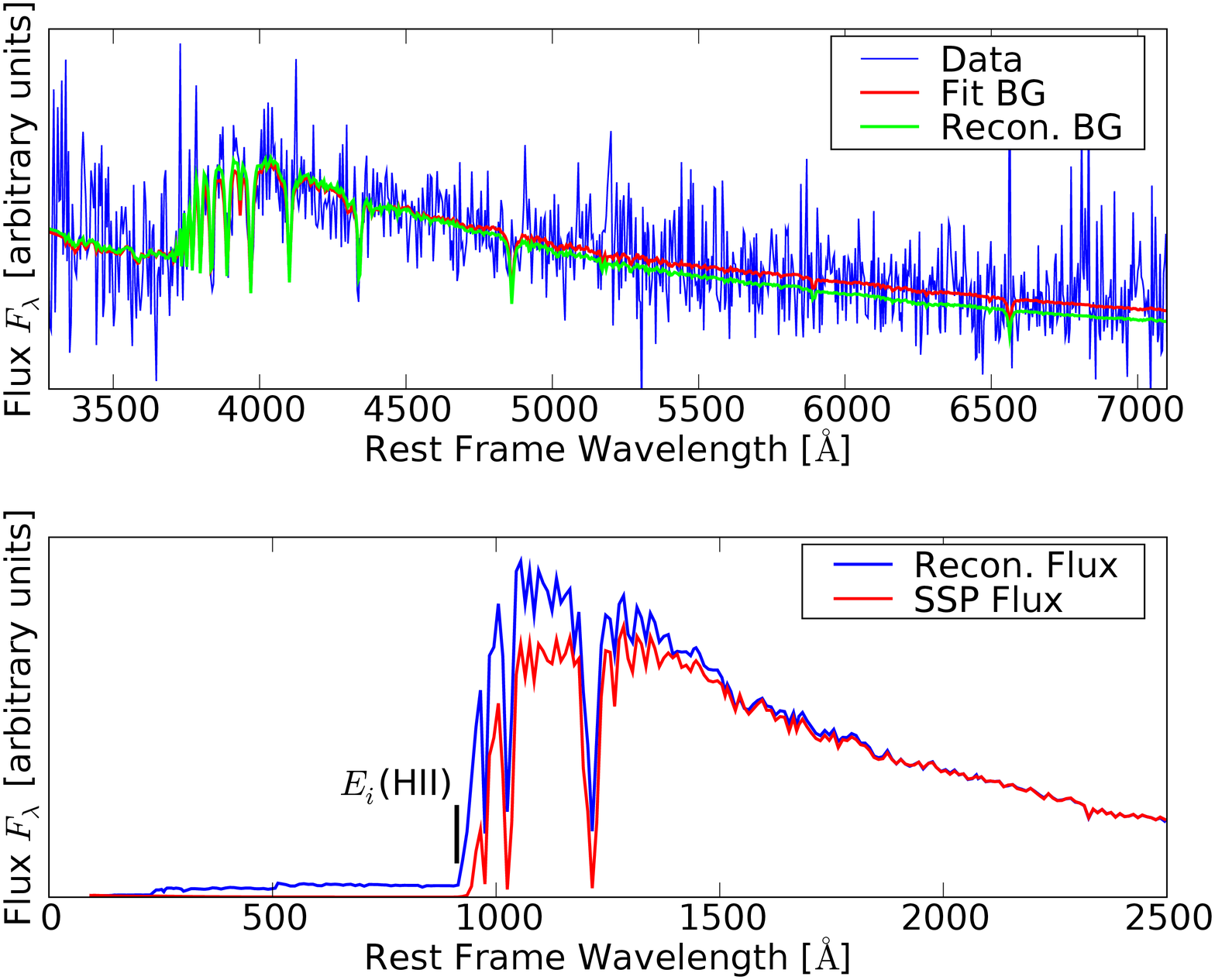}
\end{center}
\caption{\emph{Top:} Comparison of the spectrum reconstructed from the
  HOST07if age PDF(green) compared to the data (blue) and the
  background estimate from the emission-line fitting procedure
  (red). \emph{Bottom:} A comparison of the reconstructed spectrum
  (blue) and the spectrum of the SSP with age closest to the age
  estimate for HOST07if.  Note the presence of HII ionizing radiation
  in the reconstructed spectrum while the SSP (as expected) shows
  none.}
\label{fig:recon_spec}
\end{figure}

To investigate the amount of old stars in HOST07if, we begin by
reconstructing the spectrum derived by convolving the burst 2D
age-metallicity PDF with the BC03 SSPs. This ``reconstructed'' stellar
spectrum is plotted in the top panel of Figure~\ref{fig:recon_spec}
along with the data and stellar background fit from
\S\ref{sec:host_metal}.  Remarkably, information from the Balmer
absorption features and D4000 alone are enough to reconstruct much of
the HOST07if stellar background with high fidelity, especially in the
bluer wavelengths. A slight color discrepancy  of
$\Delta(g-r)\approx0.11$~mag is evident here, which could be due to
dust in HOST07if (see \S\ref{sec:reddening}) or old stars (see
discussion below). In the lower panel of the same figure, we show the
ionizing flux below the Lyman limit ($\lambda = 912$~\AA) for the
reconstructed spectrum as compared to the SSP at the age closest to
our median age and metallicity closest to our spectroscopic
measurement ($Z=0.004$). We performed a simple calculation of the
H$\alpha$ flux that would result from this ionizing flux assuming 45\%
of ionizing photons eventually generate an $H\alpha$ photon
\citep{donahue95}, and found it to be within a factor of about 2 of
the measured value.  Thus, our technique not only accurately
reproduces the stellar spectrum in the optical regime, but also
independently predicts the Balmer emission strength fairly well. This
indicates that our age-matching technique is effectively reproducing
the tapering SFR in HOST07if, which may indicate we are recovering not
only the central burst time, but also some of the morphology of the
burst SFH.

\begin{figure}[t]
\begin{center}
\includegraphics[width=0.45\textwidth]{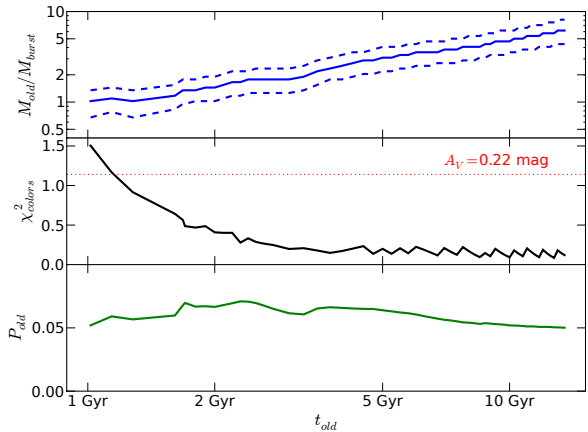}
\end{center}
\caption{\emph{Top:} Best fit value (solid blue line) and $\pm1\sigma$
  values (dashed lines) for mass of old stars at each age that when
  added to the burst spectrum produce the best matching colors ($g-r$
  and $u-g$), scaled to the mass of the burst spectrum. 
  \emph{Middle:} $\chi^2$ of colors for best-fit mass value as a
  function of age (solid black line), with the best possible $\chi^2$
  obtainable by reddening the burst spectrum (dotted red line) which
  is found at $A_V=0.22$~mag.
  \emph{Bottom:} Final probability of SN~2007if arising from old stars
  injected at a given age, derived from the product of the SN
  production likelihood with the color-matching likehood (from the
  above $\chi^2$ values).}
\label{fig:old_stars}
\end{figure}

We proceed in our investigation of possible old stars in HOST07if by
taking our reconstructed burst spectrum as being representive of the
true starburst SED. As noted above, the spectrum predicted from our
burst PDF is somewhat bluer ($\Delta(g-r)\approx0.11$ mag) than what
we observe for HOST07if, which could be a product of additional old
stars. Here we take a conservative approach and explore the
implications if the entire color excess arises from an old stellar
population. To the burst spectrum we add the SED from an additional
mass of old stars injected at a single age ranging from 1~Gyr to
13.5~Gyr. For each age, we fit for the mass of stars that minimizes
$\chi^2$ from the $g-r$ and $u-g$ colors, as well as the upper and
lower masses that produced a $\Delta\chi^2$ of 1 (i.e. $\pm1\sigma$)
from the optimum value. In Figure~\ref{fig:old_stars} (top panel) we
plot the best mass (and $\pm1\sigma$ values) of old stars (normalized
to the burst mass) as a function of age, as well as the best fitting
$\chi^2$ (middle panel). For reference in this plot, we show the best
$\chi^2$ obtainable by reddening the burst spectrum with dust (at
$R_V=3.1$), found to be $\chi^2=1.14$ at $A_V=0.22$~mag. At all ages,
the spectral features (D4000, H$\delta_A$, H$\gamma_A$) of the
old+burst spectrum differed from the observed values in HOST07if by
much less than their measured uncertainties, justifying our approach
of examining the starburst and old stellar populations separately.

This test illustrates the aforementioned fact that young bursts of
star formation tend to obscure older stellar populations. Half or more
of the stars in HOST07if could indeed come from older stars and still
be consistent with the observed spectral indices and colors, and we
can currently only disfavor old stars in HOST07if by making
assumptions about the form of its SFH. However, our age measurement
technique showed that old stars alone are inconsistent with the
observed spectral features of HOST07if, and a significant amount of
young stars dominates the galaxy spectrum. Further observational
constraints on old stars in HOST07if must wait for additional data,
such as deep imaging in the near infrared.

An old stellar population can be the source of SN~2007if only if its
reservoir of potential progenitor systems has not been
exhausted. Making the simplest assumptions -- that the original
reservoir of progenitor systems is proportional to initial stellar
mass, $M_{old}$, and that the delay time distribution (DTD) of
SN~20007if-like events is not an increasing function of the delay,
$t_{old}$, for populations older than $\sim$100~Myr -- we can define a
maximum relative rate today, given by
$(M_{old}/t_{old})/(M_{burst}/t_{burst})$, arising from any 
ancient burst of star formation. Coupling this relative rate with the
best-fit value of $M_{old}$ for each age, and scaling by the $\chi^2$
probability from the color matching, we derive the total probability
of SN~2007if having been born from an old stellar population as a
function of age. The results are shown in the bottom panel of
Figure~\ref{fig:old_stars}, where one sees that the likelihood of
SN~2007if arising from an older stellar population never exceeds about
7\% (note that this could be even lower if there is some reddening due
to dust). We therefore conclude there is a high likelihood that
SN~2007if was born in the recent burst of stars whose age was
constrained in \S\ref{sec:sfh}.  We note that mathematically this
consumption-timescale constraint gives the same relative factor as for
old stellar populations distributed equally at multiple ages -- where
consistency requires a fixed DTD normalization across ages -- and
then assuming a $t^{-1}$ power law for the DTD. This case is of
particular interest because a $t^{-1}$ power law is similar to the DTD
observed for normal 
\sneia\ \citep{maoz10,kbar10}, and expected in most DD models.

\subsection{HOST07if Stellar Mass}
The SPS models used above to constrain the luminosity-weighted
age of the HOST07if stellar population can also be used to constrain
the HOST07if mass-to-light ratio. Though spectral indices can in
principle be used to constrain the mass-to-light ratio (e.g. GB09),
the S/N of our spectral indices results in a large uncertainty
($\sim1$~dex) in the index-based mass-to-light ratio. Instead, a much
tighter constraint can be obtained using optical color. We thus
compare the $g-r$ color of HOST07if to that of our SFH models, as
this color was shown by GB09 to be a good color for constraining
mass-to-light ratios.

Which SFH models are appropriate for determining mass-to-light ratios
is a deeper question than can be addressed here. Instead we follow the
prescription generally favored in the literature, which is to use
exponentially-declining SFHs similar to those of \citet{kauff03a} and
GB09. Though the SFHs of dwarf galaxies such as HOST07if are likely to
be bursty, a long period of intermittent burst of SF can be
well-approximated by a continuous SFH. We thus use the
aforementioned suite of model spectra built following the presciptions
of GB09 to constrain the HOST07if mass-to-light ratio using $g-r$
color in the following way. Each model galaxy SED is normalized to
$M=1M_\odot$, and we measure the $g$-band luminosity for each
template. Using color-based $\chi^2$ weights, we measure the weighted
mean stellar mass of a burst of unit $g$-band luminosity as:
\begin{equation}
  \left<M\right> = \frac{\sum_i w_iM_i}{\sum_i w_i}
\end{equation}
and its uncertainty:
\begin{equation}
  \sigma_M = \left(\left<M^2\right>-\left<M\right>^2\right)^{1/2}
\end{equation}
where the weight $w_j = \exp{-\chi^2_j/2}$ for each template is
computed from the template's $g-r$ color $\chi^2$ as:
\begin{equation}
   \chi^2_j = \left[
     \frac{(g-r)_{HOST07if} - (g-r)_j}{\sigma_{(g-r)_{HOST07if}}}
     \right]^2
\end{equation}
Assuming a solar $g$-band absolute magnitude of \solarglum\
\citep{bell01}, we derive a mass-to-light ratio for HOST07if of
\finalmtolgsolar. With the absolute magnitude derived in
\S\ref{sec:host_phot} and the aforementioned mass-to-light ratio,
this implies a galaxy stellar mass for HOST07if of
\finalhostmass.

As a comparison, we inspect the mass-to-light ratios for SDSS galaxies
as determined by the MPA-JHU team. We find their $M_*/L$ values to be
well represented as a linear function in both optical $g-r$ color and
more weakly in absolute magnitude $M_g$. From their data, we estimate
the HOST07if mass-to-light ratio to be \finalmtolgsdss, which is
consistent with our value within the error bars (as would be expected
since our SFH models are essentially the same). In a similar vein, we
use the color-based $M_*/L$ formulae (appropriately corrected for our
choice of IMF) from \citet{bell01} along with the color measured from
HOST07if to estimate a mass-to-light ratio of \finalmtolbell, again
consistent with our estimate.

\section{Analysis Cross-Checks}
\label{sec:systematics}
We now discuss several cross-checks we performed in order to estimate
systematic effects in our parameter estimations. The possible effects
of dust in HOST07if, systematic uncertainties in metallicity scales,
and the limitations of our particular choice of stellar population
synthesis (SPS) models will be addressed in turn.

\subsection{The Effect of Uncertain Reddening}
\label{sec:reddening}
In typical applications the Balmer decrement is used to estimate
reddening due to dust and correct emission line metallicity
diagnostics. Our detection of H$\beta$ is consistent with no reddening,
however it is of sufficiently low S/N that a large range of reddening
is allowed by this measurement. Several lines of evidence suggest that
the reddening should be low. In \citet{scalzo10} we set strong upper
limits on the column of Na~I, suggesting little enriched material is
available in the ISM for the formation of dust. Given that HOST07if is
in a post-starburst phase, the HII regions and molecular clouds
associated with this burst will have dissipated long ago, and thus it
is quite plausible that the extinction limit derived for SN~2007if is
not atypical of that for the emitting gas and stars. Because dust
requires metals to form, the expected low metallicity based on the low
luminosity of HOST07if also leads to the expectation of low
extinction. \citet{lee09} and \citet{garn10} measure the Balmer
decrement as a function of galaxy luminosity and do indeed find that
low-luminosity galaxies typically have extinction of only
$A_V\sim0.1$.

\begin{figure}[t]
\begin{center}
\includegraphics[width=0.45\textwidth]{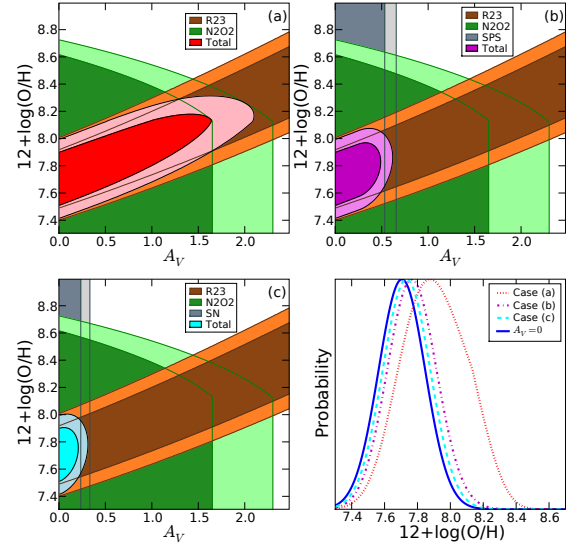}
\end{center}
\caption{\emph{Upper Left:} Two-dimensional probability distribution
  function of O/H vs. $A_V$ combining constraints from the $R_{23}$
  ratio and N2O2 ratio from emissions lines. The two filled contours
  represent the $1\sigma$ and $2\sigma$ probability levels for
  each constraint, with the red and pink contours corresponding to the
  $1\sigma$ and $2\sigma$ final combined constraints.
  \emph{Upper Right:} Same as left, but with additional constraint
  from SPS matching (see text). Magenta and fuscia contours are final
  $1\sigma$ and $2\sigma$ combined constraints.
  \emph{Lower Left:} Same as upper left, but with addition
  of SN reddening constraint from \citet{scalzo10}. Cyan and light
  blue contours are final $1\sigma$ and $2\sigma$ combined
  constraints.
  \emph{Lower Right:} Metallicity PDF (marginalizing in $A_V$) for the
  three above cases (Case a - emission line constraints only, red
  dotted line; Case b - emission line plus SPS constraints, magenta
  dash-dotted line; Case c - emission line plus SN constraints, cyan
  dashed line) as well as the simple case assuming $A_V=0$ (blue solid
  line).}
\label{fig:reddening}
\end{figure}

The emission line fluxes of HOST07if also favor low
reddening. Correction for reddening will increase $R_{23}$ and lead to
a higher predicted O/H. However, N2O2 (=[NII]$\lambda6584$/\oii)
works in the opposite sense. Indeed, at the lowest metallicities
($12+\log(O/H) < 8.1$), N2O2 is expected to saturate at primary N/O
nucleosythesis ratio of $\log(N/O)=-1.43^{0.07}_{-0.08}$
\citep{nava06}, giving N2O2$=-1.32^{+0.08}_{-0.09}$. Thus, our
non-detection of [NII]$\lambda6584$ provides an upper limit on N2O2
that can be used to constrain the amount of reddening. In the upper
left panel of Figure~\ref{fig:reddening} we show these complementary
constraints in the $A_V$--(O/H) plane.  These constraints alone
disfavor any reddening greater than $A_V\sim1.8$, and the metallicity
prediction would have been 0.21~dex higher than that derived with
fixed $A_V=0$.

The very blue color and strong Balmer absorption of the stellar
continuum in HOST07if place additional constraints on reddening. It
was noted above that the reconstructed spectrum from our
age-metallicity PDF was slightly bluer (by $\sim0.11$~mag in $g-r$ and
$0.06$~mag in $u-g$, corresponding to a best-fit extinction of
$A_V=0.22$~mag) than the observed color of HOST07if. While this could
be caused by reddening due to dust, it could also be indicative of the
presence of older stars (see \S\ref{sec:old_stars}). However, the
Balmer absorption of the HOST07if stellar continuum and its optical
colors can be combined to place an \emph{upper} limit on the amount of
reddening that is consistent with the observed color of HOST07if. This
can be understood as a disagreement between the extreme blue stellar
color implied by large reddening and the Balmer absorption strengths;
if the reddening was large and HOST07if was intrinsically much bluer,
its implied age would be younger and thus its Balmer absorption
strengths would have been shallower than observed. We can quantify
this constraint by examining the effect of reddening on the $g-r$ and
$u-g$ colors of HOST07if and the subsequent agreement with our model
spectra used in constraining the burst age. For each value of $A_V$,
we sum the probability of matching to each of the 150,000 burst
templates using the $\chi^2$ method described above with the $g-r$ and
$u-g$ colors included in the $\chi^2$. Shown in the upper right panel
of Figure~\ref{fig:reddening}, the $A_V$ PDF from this method shows a
sharp drop at $A_V\sim0.5$.  With the constraints from the stellar
features added to our PDF, the metallicity we would have measured is
only 0.08~dex higher than the $A_V=0$ value. With the $1\sigma$
reddening of $A_V\sim0.5$, our mass estimate for HOST07if would have
increased by $\sim$0.3~dex (accounting for both the luminosity and
mass-to-light ratio changes), only slightly larger than the
measurement error for that quantity.

Finally we show the result of including the SN~2007if reddening
constraint of \citet{scalzo10} as an assumed constraint on the global
host reddening. We show the resultant 2D PDF in the lower left panel of
Figure~\ref{fig:reddening}, and find that the resultant metallicity
would have been 0.02~dex higher than the $A_V=0$ value. Marginalizing
our 2D PDFs in $A_V$ gives the (O/H) PDFs for each scenario described
above, and we show these in the lower right panel. The values we
reported for each scenario represent the metallicity of maximum
likelihood for the PDFs shown. Thus, while there is some uncertainty
in the amount of reddening in HOST07if because we cannot strongly
constrain the Balmer decrement, ultimately it has little effect on our
final results, which robustly show a low metallicity. Additionally,
the spectral indices used in our age measurement are measured across
short wavelength ranges and thus are relatively insensitive to
reddening, making our age estimate also robust against possible
reddening in HOST07if.

\subsection{Metallicity Calibration}
\label{sec:metal_scales}
Strong emission-line methods such as the $R_{23}$ method
\citep[][KK04]{m91,z94} produce oxygen abundance values that are
systematically higher than those derived with the direct method by
$0.2-0.5$~dex \citep{kbg03}. In galaxies like HOST07if where \aoiii\
is not detected with sufficient S/N but strong line fluxes indicate low
metallicity, this poses a challenge for deriving the correct absolute
metallicity. Placing our metallicity estimate on the correct
absolute scale is subject to the uncertainty as to which metallicity
calibration is correct in an absolute sense.  This is a subject much
debated, and while the final scale remains undecided, \citet{ke08}
provided an excellent analysis of the discrepancies between various
scales and means of converting between them. The scatter in these
relations (0.06~dex systematic error, see analysis below) is 
smaller than the measurement errors from our spectrum.  Placement
of our measurements on a common scale with those of other \snia\ hosts
in the literature suffices for comparison purposes, and will be
employed in the discussion of \S\ref{sec:other_snia_host_metal}.

\subsection{Systematics in Stellar Population Synthesis}
Next, we consider the impact of our particular choice of SPS
models. It is a well-known problem that different stellar evolution
and population synthesis codes produce different results due to
different treatment of uncertain stages of stellar evolution,
extinction due to dust, and the IMF \citep[see
e.g.][]{cgw09}. Assessment of the full impact of these uncertainties 
is beyond the scope of this work, but we inspected the impact of
employing a \citet{salpeter} IMF instead of the \citet{chab03} IMF
used in our primary analysis. The age constraint for HOST07if remained
unchanged, as the Salpeter IMF increases the amount of low-mass stars
(relative the Chabrier IMF) which negligibly affect the spectrum of
young starbursts similar to HOST07if which are dominated by bright
massive stars. The mass-to-light ratio, however, is $\approx0.16$~dex
higher for the Salpeter IMF, again a product of the increased
proportion of low mass stars. Thus while our host mass is (as
expected) dependent on the IMF chosen, the age constraint is robust
against different IMFs.

To place our results in a more general
stellar population context, we inspected the stellar spectra catalog
of \citet{gunnstryker83} and measured the Balmer absorption strengths
in the same manner as for HOST07if. We then analyzed which single star
spectra had the closest absorption strengths to HOST07if, and found
the majority of these to be late B-type or early A-type stars. This is
consistent with the age and main-sequence turnoff mass derived for
HOST07if. Thus we find that our age measurement for the stellar
population dominating the light of HOST07if is consistent with
single-star spectra, indicating that our results are unlikely to be
strongly dependent on the choice of SPS models.

\subsection{Summary}
In summary, the possible systematic errors or biases on our
measurements of the metallicity and age of HOST07if are small compared 
to measurement errors. Our classification of HOST07if as metal-poor is
confirmed for a wide variety of assumptions about reddening by dust in
the host, and is true regardless of the metallicity calibration
chosen. Our measurement of the young age of the stellar populations in
HOST07if is not an artifact of our choice of SPS models or template
SFHs. One important subtlety to note is that our age PDF for the
stellar populations in HOST07if does not constitute a direct
measurement of the progenitor age of SN~2007if, as the SN progenitor
system was drawn from a single epoch in the SFH of its host
galaxy. Our estimate of the stellar ages of the host represents the
distribution of ages from which the progenitor was drawn, rather than
a constraint on the age of the single progenitor system. The
statistics of the host age distribution strongly favor a young age for
the progenitor system of SN~2007if. Our assertion that HOST07if is
young and metal-poor is robust, and serves as appropriate context for
considering the properties of the progenitor of SN~2007if.

\section{Discussion}
\label{sec:discussion}
In this section we discuss HOST07if in the context of previous
\snia\ host galaxy studies, as well as the implications for progenitor
scenarios for SN~2007if suggested by our data. Our assumption is that
the properties of the host galaxy stellar population are good
indicators of the properties of the progenitor system of SN~2007if.
We showed above that this argument is statistically sound, as the
recent major starburst dominates the galaxy light.  Below we will show
that our results are consistent with regions of progenitor parameter
space believed to produce \sneia, and our results thus provide
important constraints on what portions of that parameter space are
likely to produce \superc\ \snia.

\subsection{Metallicity of SN~2007if host - Comparison to other \snia\
hosts}
\label{sec:other_snia_host_metal}
Metallicity is a key parameter affecting the evolution of \snia\
progenitors. In the SD scenario \citep{hkn08}, accretion is stabilized
by a strong wind from the WD whose strength is driven by Fe opacity.
Lower metallicity decreases the allowable regions of WD mass - orbital
period parameter space in which the wind is strong enough to stabilize
the accretion \citep{kn09}. In general, metallicity will affect the
relation between initial main-sequence mass and final WD mass, as well
as the time to evolve off the main sequence \citep{umeda99a}.  For WDs
of the same mass, a lower metallicity produces a slightly lower C/O
ratio (a product of the aforementioned evolution time effect), which
has been proposed as a possible source of the diversity in \snia\
brightnesses \citep{umeda99b}.

Placing our metallicity measurement in the context of previously
published spectroscopic \snia\ host metallicities requires using a
common scale, as different metallicity calibrations produce
significantly different results \citep[see the excellent discussion
in][]{ke08}.  To our knowledge, the lowest spectroscopic \snia\ host
metallicities to-date are those of SN~1972E at $12+\log(O/H)=8.14$
\citep{hamuy00}, and SN~2004hw at $12+\log(O/H)=8.23$
\citep{prieto08}. The original metallicity of the host of SN~1972E is
drawn from \citet{kobul99}, who use the ``direct'' method to measure
the oxygen abundance.  As noted above, the ``direct'' method values
are typically lower than strong line values by at least 0.2
dex. Therefore we collected the galaxy emission line fluxes from
\citet{kobul99} and measured its abundance using the KK04 technique
employed for HOST07if, finding $12+\log(O/H)_\mathrm{72E,KK04}=8.35
\pm 0.03$. The metallicities of \citet{prieto08} come from \snia\
hosts in the T04 sample, where metallicities where derived in a
Bayesian manner by comparing emission line fluxes to photoionization
models of \citet{cl01}. While the means to reproduce their metallicity
analysis are not available, the absorption-corrected emission line
fluxes are available from the MPA-JHU group. Using the fluxes for the
host of SN~2004hw, we find $12+\log(O/H)_\mathrm{04hw,KK04}=8.24 \pm
0.03$. After placing all these metallicities on a common scale, our
value of \finalmetalkk\ for the metallicity of HOST07if is $\approx
2\sigma$ lower than the lowest metallicity from these previous
samples, and far below the metallicities of typical \snia\ host
galaxies.

Interpretation of the metallicity of HOST07if on an absolute scale is
subject to the inter-calibration issues described above. The T04 scale
is a popular one in the literature, as the mass-metallicity relation
they derive is often invoked to use host mass as a proxy for
metallicity. As stated above, the algorithm for this scale is not
accessible, but we can convert our values to this scale using the
conversion formulae of \citet{ke08}. Doing so yields a metallicity for
HOST07if of \finalmetaltrem, and for the host of
SN~1972E yields $12+\log(O/H)_\mathrm{72E,T04}=8.22$, while the value
for SN~2004hw of $12+\log(O/H)_\mathrm{04hw,T04}=8.23$ was derived in 
the T04 data set. On this scale HOST07if is nearly $3\sigma$ lower
metallicity than the other \snia\ hosts.

We can place the metallicity of HOST07if on a solar abundance
scale by comparing our measurements of the oxygen abundance to the
solar value of $12+\log(O/H)_\odot=8.86$ \citep{delahaye10}. On the
KK04 scale, HOST07if has metallicity $Z_\mathrm{KK04} \approx
Z_\odot/5$, while on the T04 scale it has $Z_\mathrm{T04} \approx
Z_\odot/9$. The T04 value is perhaps in better agreement with the
stellar metallicity preferred by our template matching (see
\S\ref{sec:sfh}) where 40\% of the probability is in
the $Z=0.004$ ($Z_\odot/5$) track and 25\% in the $Z=0.0004$
($Z_\odot/50$), though the coarse metallicity binning of the BC03
models makes this difficult to quantify precisely. While the absolute
scale is somewhat uncertain, the metallicity of HOST07if is
significantly sub-solar on several reasonable metallicity scales.

Finally we note that the gas-phase metallicity measured for HOST07if
at the present epoch may be higher than the metallicity at the time of
the birth of the progenitor of SN~2007if, presumably during the major
starburst \finalageapprox\ before the SN. The massive stars formed
during that burst have exploded as core collapse SNe and enriched the
ISM of HOST07if with their ejecta. \citet{sa09} found that dwarf
galaxies with bursty SFHs showed gas-phase metallicities enriched by
$\approx0.35$~dex (as compared to stellar metallicities) during the
periods between bursts of star-formation. Thus the metallicity of the
progenitor system of SN~2007if could possibly be even lower than the
extremely low gas-phase metallicity measured for HOST07if at the time
of the SN itself.

\subsection{\sneia\ in low luminosity hosts}
SN~2007if is part of a large and continually growing list of unusual
\sneia\ discovered in low-luminosity host galaxies. Though
low-luminosity galaxies have a higher number density than
high-luminosity galaxies due to the steep faint-end slope of the
galaxy luminosity function \citep[e.g.][]{schecter76}, high-luminosity
galaxies retain the majority of stellar mass and thus are likely to
produce the large majority of supernovae. Despite this fact, the
number of supernovae in low-luminosity hosts is now significant, and
includes a number of peculiar SNe such as SN~2007if.

The SN~2002cx-like
supernova SN~2008ha \citep{foley09} was found in a faint ($M_B =
-18.2$ for $h=0.7$) irregular galaxy.  SN~2002ic
\citep{mwv02,hamuy03,mwv04} and SN~2005gj \citep{aldering06,prieto07},
both of which demonstrated
features consistent with interaction with circumstellar material,
were found  in low-luminosity hosts \citep[as-yet undetected for
SN~2002ic and $M_B = -17.4$ for SN~2005gj;][]{aldering06}. At the
most extreme, SN~1999aw was found in a host galaxy of brightness $M_B
= -11.9 \pm 0.2$ \citep{strolger02}.  The prototype of the possible
\superc\ class, SN~2003fg, was discovered in a low luminosity galaxy
whose mass was estimated at $\log(M_*/M_\odot) = 8.93$
\citep{howell06}, though it is possible this is a tidal feature of a
larger morphologically-disturbed galaxy nearby.

While the prevalence of unusual \sneia\ in low-luminosity galaxies is
intriguing, it is by no means a one-to-one relationship.  Most of the
SN~2002cx-like host galaxies are spirals of moderate stellar mass
\citep{foley09}, and other \superc\ candidates have been found in more
massive galaxies.  SN~2006gz was found in a bright Scd galaxy
\citep{hicken07}, and SN~2009dc appeared to be located in a massive S0
galaxy but may be associated with a nearby blue companion at the same
redshift which may be interacting with the fiducial host of SN~2009dc
\citep{silverman10,tauben10}. 

There have also been relatively normal \sneia\ in low
luminosity galaxies.  The host galaxy of SN~2006an has an extremely
low luminosity ($M_g = -15.3$, SDSS) and stellar mass
\citep[$\log(M_*/M_\odot) = 7.7$,][]{kelly10} but was 
matched spectroscopically to the normal \snia\ SN~1994D
\citep{quimby06an}. The Catalina Real-Time Transient Survey
\citep{crts} discovered SN~2008hp in a very faint ($M_g = -12.7$)
host galaxy, but matched it spectroscopically to a normal \snia\
\citep{drake08hp}. Additionally, the \snf\ discovered a number of
relatively normal \sneia\ in low luminosity galaxies (Childress \etal,
in preparation). 

To summarize, we note that low luminosity \snia\ hosts do not
exclusively produce unusual \sneia, but there appears to be a higher
frequency of these peculiar \sneia, including SN~2007if, in lower
luminosity hosts.

\subsection{Host Age Constraint - 
Implications for SN~2007if Progenitor Scenarios}
A consistent picture for the progenitor of any supernova should be
able to explain not only the energetics of the explosion itself, but
also the rates and timescales of such events.  For normal \sneia, the
correlation of SN rates with host galaxy mass and star-formation rate
\citep{mannucci05,sullivan06} indicated the likelihood of two
progenitor components \citep[the ``A+B'' model][]{scan05} with
different time scales.  This is most directly encapsulated in
the \snia\ delay time distribution \citep[DTD;][]{mannucci06}.  While
the DTD of \sneia\ is still debated \citep[see e.g.][]{mennekens10},
the predictions of various scenarios for normal \sneia\ serve as a
useful baseline for placing our age constraint for HOST07if in the
context of progenitor scenarios for SN~2007if.

Though the presence of \sneia\ in elliptical galaxies and the decline
of the \snia\ rate at high redshift argue for progenitors with long
delay times \citep{strolger04,strolger05}, the correlation of \sneia\
with star-formation indicates the need for short-lived \snia\
progenitors \citep{aubourg08} with delay times of order a few hundred
Myr.  Such short timescales have indeed been obtained in models of SD
progenitor scenarios \citep[e.g.][their WD+MS channel]{hkn08}, and DD
scenarios \citep[e.g.][]{ruiter09}.  

As an example, \citet{mennekens10} describe a particular DD channel
(dubbed the ``CE'' channel) in which two stars, with initially large
separation and orbital period of several hundred days, undergo two
common-envelope phases at the end of the main sequence lifetime of the
more massive star. Following the MS evolution and the CE episodes, the
orbital period of the system is reduced to a few hundred seconds and
rapidly decays by gravitational radiation over a few hundred
kyr. Finally the two WDs merge after a total period 
of order a few hundred Myr from the initial birth of the stars.  This
binary evolution channel has delay times consistent with our age
estimate for the stellar population of HOST07if.

\citet{liu10} proposed a stellar evolution channel for \superc\
\sneia\ involving a CO-WD primary and He secondary.  This system is
born from a binary with initial masses of $M_1=7.5M_\odot$ and
$M_2=4.0M_\odot$ (at solar metallicity) that undergoes rapid rotation
and explodes as a \superc\ \snia\ with a delay time of approximately
$t_{super-Ch} \approx 65$~Myr.  Though the initial masses and
timescales would be different at the sub-solar metallicity of
HOST07if, the timescale of this scenario is roughly the same order of
magnitude as our age constraint from the host spectrum.

\citet{bn11} proposed a new binary evolution scenario which
could lead to an \snia\ through the DD channel.  In their ``single
CE'' scenario, two stars of very similar mass ($M_1/M_2 > 0.95$) fill
their Roche lobes almost simultaneously, leading to a common envelope
episode that brings the remnant WDs to a much tighter orbital
separation followed by the standard DD merger as a result of orbital
energy dissipation due to gravitational radiation losses. Their
scenario manifests a large range of timescales, from less than 
100~Myr to greater than a Hubble time, which allows for the timescale
that we estimate for the age of HOST07if. 
Indeed, a non-negligible fraction of the short timescale ($\log(t)
\leq 8.2$) realizations of this scenario show a total WD system mass
in the range $2.1~M_\odot \leq M_{WD,tot} \leq 2.3~M_\odot$
(L. Nelson, private communication), in line with to the total system
mass estimate we derived for SN~2007if in \citet{scalzo10}.

Short timescales similar to the age of HOST07if are allowed in some SD
scenarios [e.g. the ``WD+MS'' channel of\citet{hkn08}, see also
  \citet{hanpods04}, \citet{greggio05} and references therein], but
are especially common in DD scenarios
\citep{yunglivio00,greggio05,ruiter09,mennekens10}.
While our age constraint does not definitely establish whether one of
the traditional \snia\ progenitor scenarios or a new scenario is more
favored for SN~2007if, our determination that SN~2007if was likely
born from a young stellar population disfavors some scenarios,
such as the WD+RG channel of the SD scenario from \citet{hkn08} in
which the WD accretes matter from a red giant companion, or the
``RLOF'' channel of the DD scenario described by \citet{mennekens10}
in which early mass transfer in the binary proceeds by slow Roche lobe
overflow (RLOF) and requires a significantly longer delay time than
the age we measure for HOST07if.

Another interesting consequence of our age constraint is the resultant
progenitor WD mass constraint for SN~2007if. If we assume SN~2007if
originated from the merger of two WDs born in the dominant HOST07if
starburst that have evolved off the main sequence prior to merger,
we can use models connecting initial MS and final WD mass to derive a
crude lower limit for the total system mass prior to \snia\
explosion. Using the models of \citet[][see their Fig. 6]{umeda99a} at
$Z=0.004$, we roughly estimate that a \finalmstoplain\ main
sequence star (corresponding to the MS turnoff mass derived above for
HOST07if) would produce a \finalwdmass\ white dwarf.
Thus in this toy model SN~2007if should have originated from the
merger of two WDs whose total mass can be no less than
\finalsystemmass, clearly in excess of $M_{Ch}$. There must some
dynamical orbital decay time for a double WD merger, so this
approximation should be considered an extreme lower limit. Though the
evolution of post-MS stars in binary systems is far more complicated
than the single star evolutionary scenarios of \citet{umeda99a}, these
models provide a good approximate scale of the available C/O material
at the time of WD merger.  Thus, our age estimate for HOST07if implies
that even if stars just leaving the main sequence in HOST07if merge
immediately, their mass must exceed the Chandrasekhar mass by a fair
margin, reinforcing the model of SN~2007if as a \superc\ \snia\ we
derived in \citet{scalzo10}.

\section{Conclusions}
\label{sec:conclusions}
We have presented Keck photometry and spectroscopy of the faint host
galaxy of the \superc\ \snia\ SN~2007if.  HOST07if has very low
stellar mass (\finalhostmass), and has the lowest-reported
spectroscopically-measured metallicity (\finalmetalkk\ or
\finalmetaltrem) of any \snia\ host galaxy. We used the Balmer
absorption line strengths in conjunction with the 4000\AA\ break to
constrain the age of the dominant starburst in the galaxy to be
\finalhostage, corresponding to a main-sequence turn-off mass of
\finalhostmsto.

This host galaxy is an ideal system for measuring SN progenitor
properties.  Dwarf galaxies such as HOST07if typically have a
well-mixed ISM, lacking the large-scale abundance gradients found in
larger galaxies.  Like other low-mass dwarf galaxies, HOST07if shows
indications of a bursty star-formation history, as its recent
star-formation is dominated by the large starburst approximately
\finalageapprox\ in its past which presumably gave birth to the
progenitor system of SN~2007if.
We note, however, that bright recent starbursts are efficient at
obscuring the light of older stellar populations, and HOST07if could
possibly have a significant amount of mass in older stars
(c.f. \S\ref{sec:old_stars}). However, we also showed that with the
decreased probability of SN~2007if arising from progressively older
stars, the allowable amount of old stars in HOST07if leaves only a
small probability that the SN was not born in the most recent
starburst.
Our constraints on the age and metallicity of the host of SN~2007if do
not constitute direct constraints on the properties of its progenitor,
but rather characterize the distribution of stars from which its
progenitor was drawn.
Nonetheless, the low metallicity and young stellar age of HOST07if are
robust measurements (c.f. \S\ref{sec:systematics}), and strengthen
our interpretation that the properties of HOST07if are good indicators
of the properties of the SN~2007if progenitor itself.

Our results provide key properties that should be reproduced by any 
proposed progenitor scenarios for SN~2007if.  The low host metallicity
can be used as input to stellar evolutionary tracks chosen for
progenitor modeling, and will be particularly important in the mass
loss stages of the progenitor.  The relatively short timescale for the
explosion of SN~2007if provides constraints on the binary evolution of
the progenitor system. While development of a consistent progenitor
scenario for \superc\ \sneia\ is beyond the scope of this work, we
have shown that a key member of this subclass, SN~2007if, is very
likely to have originated from a low-metallicity young
progenitor. Future inspection of the hosts of other \superc\ \sneia\
will be critical for assessing the frequency of these characteristics
for \superc\ \snia\ progenitors.

\vskip11pt

\scriptsize
Acknowledgments: The authors would like to thank the excellent
technical and scientific staff at Keck Observatory.  The data
presented herein were obtained at the W. M. Keck Observatory, which is
operated as a scientific partnership among the California Institute of
Technology, the University of California, and the National Aeronautics
and Space Administration; the Observatory was made possible by the
generous financial support of the W. M. Keck Foundation. We wish to
recognize and acknowledge the very significant cultural role and
reverence that the summit of Mauna Kea has always had within the
indigenous Hawaiian community, and we are extremely grateful for the
opportunity to conduct observations from this mountain. We thank Lorne
Nelson and Eric Blais for insightful discussions about their \snia\
progenitor work. We would also like to thank the anonymous referee for
very constructive comments which helped improve the quality of this
paper.

This work was supported by the Director, Office of Science, Office of
High Energy Physics, of the U.S. Department of Energy under Contract No.
DE-AC02-05CH11231; by a grant from the Gordon \& Betty Moore Foundation;
and in France by support from CNRS/IN2P3, CNRS/INSU, and PNC.
CW acknowledges support from the National Natural Science Foundation
of China grant 10903010.
This research used resources of the National Energy Research Scientific
Computing Center, which is supported by the Director, Office of Science,
Office of Advanced Scientific Computing Research, of the U.S. Department
of Energy under Contract No. DE-AC02-05CH11231.  We thank them for a generous
allocation of storage and computing time.
HPWREN is funded by National Science Foundation Grant Number ANI-0087344,
and the University of California, San Diego.

Some data used in this paper were obtained from the Sloan Digital Sky
Survey (SDSS).  Funding for the SDSS and SDSS-II has been provided by
the Alfred P. Sloan Foundation, the Participating Institutions, the
National Science Foundation, the U.S. Department of Energy, the
National Aeronautics and Space Administration, the Japanese
Monbukagakusho, the Max Planck Society, and the Higher Education
Funding Council for England. The SDSS Web Site is
http://www.sdss.org/. 



\begin{thebibliography}{}
\bibitem[Abazajian et al.(2009)]{abaz09} 
Abazajian, K., et al., 2009, \apjs, 182, 543

\bibitem[Akerlof et al.(2007)]{akerlof07}
Akerlof, C. et al., 2007, CBET, 1059, 1

\bibitem[Aldering et al.(2002)]{ald02} 
Aldering, G., et al., 2002, SPIE, 4836, 61

\bibitem[Aldering et al.(2006)]{aldering06} 
Aldering, G., et al., 2006, \apj, 650, 510

\bibitem[Aubourg et al.(2008)]{aubourg08} 
Aubourg, E., et al., 2008, \aap, 492, 631

\bibitem[Balogh et al.(1999)]{D4000}
Balogh, M., et al., 1999, \apj, 527, 54

\bibitem[Barbary et al.(2010)]{kbar10}
Barbary, K. et al., 2010, arxiv:1010.5786

\bibitem[Bell \& de Jong(2001)]{bell01}
Bell, E., \& de Jong, R., 2001, \apjs, 550, 212

\bibitem[Bertin \& Arnouts(1996)]{sextractor}
Bertin, E., \& Arnouts, S. 1996, \aaps, 117, 393

\bibitem[Bertin et al.(2002)]{bertin02} 
Bertin, E., Mellier, Y., Radovich, M., Missonnier, G., Didelon, P., 
\& Morin, B.\ 2002, 
Astronomical Data Analysis Software and Systems XI, 281, 228 

\bibitem[Bertin(2006)]{bertin06} 
Bertin, E.\ 2006, Astronomical 
Data Analysis Software and Systems XV, 351, 112 

\bibitem[Blais \& Nelson(2011)]{bn11} 
Blais, E., \& Nelson, L.\ 2011, 
Bulletin of the American Astronomical Society, 43, \#324.03 

\bibitem[Bruzual \& Charlot(2003)]{bc03}
Bruzual, G., \& Charlot, S., 2003, \mnras, 344, 1000

\bibitem[Cardelli, Clayton, \& Mathis(1989)]{ccm}
Cardelli, J., Clayton, G. \& Mathis, J. 1989, \apj, 345, 245

\bibitem[Cardiel et al.(1998)]{cardiel98}
Cardiel, N., et al., 1998, \aaps, 127, 597

\bibitem[Chabrier(2003)]{chab03}
Chabrier, G., 2003, \pasp, 115, 763

\bibitem[Charlot \& Longhetti(2001)]{cl01}
Charlot, S., \& Longhetti, M., 2001, \mnras, 323, 887

\bibitem[Conroy, Gunn, \& White(2009)]{cgw09}
Conroy, C., Gunn, J. \& White, M. 2009, \apj, 699, 486

\bibitem[Delahaye et al.(2010)]{delahaye10}
Delahaye, F., et al., 2010, arXiv:1005.0423 

\bibitem[Donahue, Aldering, \& Stocke(1995)]{donahue95}
Donahue, M., Aldering, G. \& Stocke, J. 1995, \apj, 450, L45

\bibitem[Drake et al.(2009)]{crts}
Drake, A., et al., 2009, \apj, 696, 870

\bibitem[Drake et al.(2008)]{drake08hp}
Drake, A., et al., 2008, CBET, 1589, 1

\bibitem[Foley et al.(2009)]{foley09} 
Foley, R., et al., 2009, \aj, 138, 376

\bibitem[Gallazzi \& Bell(2009)]{gb09}
Gallazzi, A., \& Bell, E., 2009 (GB09), \apjs, 185, 253

\bibitem[Gallazzi et al.(2005)]{gallazzi05}
Gallazzi, A.. et al., 2005, \mnras, 362, 41

\bibitem[Garn \& Best(2010)]{garn10}
Garn, T. \& Best, P., 2010, \mnras, 409, 421

\bibitem[Graves \& Schiavon(2008)]{ez_ages}
Graves, G., \& Schiavon, R., 2008, \apjs, 177, 446

\bibitem[Greggio(2005)]{greggio05}
Greggio, L., 2005, \aap, 441, 1055

\bibitem[Gunn \& Stryker(1983)]{gunnstryker83}
Gunn, J., \& Stryker, L., 1983, \apjs, 52, 121

\bibitem[Hachisu, Kato, \& Nomoto(2008)]{hkn08}
Hachisu, I., Kato, M., \& Nomoto, K., 2008, \apj, 679, 1390

\bibitem[Hamuy et al.(2000)]{hamuy00}
Hamuy, M. et al., 2000, \aj, 120, 1479

\bibitem[Hamuy et al.(2003)]{hamuy03}
Hamuy, M. et al., 2003, \nat, 424, 651

\bibitem[Han \& Podsiadlowski(2004)]{hanpods04}
Han, Z., \& Podsiadlowski, P., 2004, \mnras, 350, 1301

\bibitem[Hanuschik(2003)]{uves}
Hanuschik, R. W., 2003, \aap, 407, 1157

\bibitem[Hicken et al.(2007)]{hicken07}
Hicken, M. et al., 2007, \apj, 669, L17

\bibitem[Hinshaw et al.(2009)]{hinshaw09} 
Hinshaw, G., et al., 2009, \apjs, 180, 225

\bibitem[Howell et al.(2006)]{howell06}
Howell, D. A. et al., 2006, \nat, 443, 308

\bibitem[Iben \& Tutukov(1984)]{iben84}
Iben, I., \& Tutukov, A., 1984, \apjs, 54, 335

\bibitem[Kauffmann et al.(2003a)]{kauff03a}
Kauffmann, G.. et al., 2003a, \mnras, 341, 33

\bibitem[Kelly et al.(2010)]{kelly10}
Kelly, P. et al., 2010, \apj, 715, 743

\bibitem[Kennicutt(1998)]{kenn98}
Kennicutt, R., 1998, \araa, 36, 189

\bibitem[Kennicutt, Bresolin \& Garnett(2003)]{kbg03}
Kennicutt, R., Bresolin, F., \& Garnett, D., 2003, \apj, 591, 801

\bibitem[Kewley \& Dopita (2002)]{kd02}
Kewley, L., \& Dopita, M., 2002, ApJS, 142, 35

\bibitem[Kewley \& Ellison(2008)]{ke08}
Kewley, L., \& Ellison, S., 2008, \apj, 681, 1183

\bibitem[Kobayashi \& Nomoto(2009)]{kn09}
Kobayashi, C., \& Nomoto, K., 2009, \apj, 707, 1466

\bibitem[Kobulnicky \& Kewley(2004)]{kk04}
Kobulnicky, H., \& Kewley, L., 2004, \apj, 617, 240

\bibitem[Kobulnicky et al.(1999)]{kobul99}
Kobulnick, H. et al., 1999, \apj, 514, 544

\bibitem[Lee et al.(2009)]{lee09}
Lee, J., et al., 2009, \apj, 706, 599

\bibitem[Leibundgut (2000)]{leibundgut00}
Leibundgut, B., 2000, \araa, 10, 179

\bibitem[Liu et al.(2010)]{liu10}
Liu, W.-M.. et al., 2010, \aap, 523, 3

\bibitem[Mannucci et al.(2005)]{mannucci05}
Mannucci, F., et al., 2005, \aap, 433, 807

\bibitem[Mannucci et al.(2006)]{mannucci06}
Mannucci, F., et al., 2006, \mnras, 370, 773

\bibitem[Maoz(2010)]{maoz10}
Maoz, D., 2010, AIPC, 1314, 223

\bibitem[McGaugh(1991)]{m91}
McGaugh, S., 1991, \apj, 380, 140

\bibitem[Mennekens et al.(2010)]{mennekens10}
Mennekens, N., et al., 2010, \aap, 515, 89

\bibitem[Mink(2006)]{mink06}
Mink, D.\ 2006, Astronomical Data 
Analysis Software and Systems XV, 351, 204 

\bibitem[Nava et al.(2006)]{nava06} 
Nava, A., et al., 2006, \apj, 645, 1076

\bibitem[Nomoto et al.(1995)]{nomoto95}
Nomoto, K., et al., 1995, ASPC, 72, 164

\bibitem[Nugent(2007)]{nugent07}
Nugent, P., 2007, ATEL \#1213

\bibitem[Nugent, Kim, \& Perlmutter(2002)]{nkp02}
Nugent, P., Kim, A., \& Perlmutter, S., 2002, \pasp, 114, 803

\bibitem[Oke et al.(1995)]{oke95}
Oke, J. B., et al., 1995, \pasp, 107, 375

\bibitem[Osterbrock \& Ferland(2006)]{agn2} 
Osterbrock, D.~E., \& Ferland, G.~J.\ 2006, Astrophysics of gaseous
nebulae and active galactic nuclei, 2nd.~ed.~by D.E.~Osterbrock and
G.J.~Ferland.~Sausalito, CA: University Science Books, 2006.


\bibitem[Peeples et al.(2008)]{peeples08}
Peeples, M., et al., 2008, \apj, 685, 904

\bibitem[Perlmutter et al.(1999)]{42sne}
Perlmutter, S., et al., 1999, \apj, 517, 565

\bibitem[Phillips et al.(2006)]{keckadc} 
Phillips, A., et al., 2007, SPIE, 6269, 56

\bibitem[Piro(2008)]{piro08}
Piro, A., 2008, \apj, 679, 616

\bibitem[Prieto et al.(2007)]{prieto07}
Prieto, J., et al., 2007, arxiv:0706.4088

\bibitem[Prieto et al.(2008)]{prieto08}
Prieto, J., Stanek, K., \& Beacom, J., 2008, \apj, 673, 999

\bibitem[Quimby et al.(2006)]{quimby06an}
Quimby, R., et al., 2006, CBET, 413, 1

\bibitem[Raskin et al.(2010)]{raskin10}
Raskin, C., et al., 2010, \apj, 724, 111

\bibitem[Riess et al.(1998)]{riess98} 
Riess, A., et al., 1998, \aj, 116, 1009

\bibitem[Ruiter et al.(2009)]{ruiter09}
Ruiter, A., et al., 2009, \apj, 699, 2026

\bibitem[Salpeter(1955)]{salpeter}
Salpeter, E. E., 1955, \apj, 121, 161

\bibitem[S\'anchez Almeida et al.(2008)]{sa08}
S\'anchez Almeida, J., et al., 2008, \apj, 685, 194

\bibitem[S\'anchez Almeida et al.(2009)]{sa09}
S\'anchez Almeida, J., et al., 2009, \apj, 698, 1497

\bibitem[Scalzo et al.(2010)]{scalzo10} 
Scalzo, R., et al., 2010, \apj, 713, 1073

\bibitem[Scannapieco \& Bildsten(2005)]{scan05}
Scannapieco, E., \& Bildsten, L., 2005, \apjl, 629, 85

\bibitem[Schecter(1976)]{schecter76}
Schecter, P., 1976, \apj, 203, 297

\bibitem[Schlegel, Finkbeiner \& Davis(1998)]{sfd98}
Schlegel, D. J., Finkbeiner, D. P. \& Davis, M., 1998, \apj, 500, 525

\bibitem[Searle \& Sargent(1972)]{ss72}
Searle, L., \& Sargent, W., 1972, \apj, 173, 25

\bibitem[Silverman et al.(2011)]{silverman10}
Silverman, J. M. et al., 2011, \mnras, 410, 585

\bibitem[Skrutskie et al.(2006)]{twomass} 
Skrutskie, M., et al., 2006, \aj, 131, 1163

\bibitem[Stoughton et al.(2002)]{stoughton02} 
Stoughton, C., et al., 2002, \aj, 123, 485

\bibitem[Strolger et al.(2002)]{strolger02}
Strolger, L., et al., 2002, \aj, 124, 2905

\bibitem[Strolger et al.(2004)]{strolger04}
Strolger, L., et al., 2004, \apj, 613, 200

\bibitem[Strolger et al.(2005)]{strolger05}
Strolger, L., et al., 2005, \apj, 635, 1370

\bibitem[Sullivan et al.(2006)]{sullivan06}
Sullivan, M., et al., 2006, \apj, 648, 868

\bibitem[Tanaka et al.(2010)]{tanaka10} 
Tanaka, M., et al., 2010, \apj, 714, 1209

\bibitem[Taubenberger et al.(2010)]{tauben10}
Taubenberger,S. et al., 2010, arXiv:1011:5665

\bibitem[Tremonti et al.(2004)]{trem04}
Tremonti, C., et al., 2004, \apj, 613, 898

\bibitem[Umeda et al.(1999a)]{umeda99a}
Umeda, H., et al., 1999a, \apj, 513, 861 

\bibitem[Umeda et al.(1999b)]{umeda99b} 
Umeda, H., et al., 1999b, \apjl, 522, L43 

\bibitem[van Dokkum(2001)]{vandokkum01}
van Dokkum, P., 2001, \pasp, 113, 1420

\bibitem[Vazdekis et al.(2010)]{vazdekis10}
Vazdekis, A., et al., 2010, \mnras, 404, 1639

\bibitem[Whelan \& Iben(1973)]{whelan73}
Whelan, J., \& Iben, I., 1973, \apj, 186, 1007

\bibitem[Wood-Vasey et al.(2002)]{mwv02}
Wood-Vasey, M., et al., 2002, IAUC, 7842, 1

\bibitem[Wood-Vasey, Wang, \& Aldering(2004)]{mwv04}
Wood-Vasey, M., Wang, L. \& Aldering, G. 2004, \apj, 616, 339

\bibitem[Worthey et al.(1994)]{worthey94}
Worthey, G., et al., 1994, \apjs, 94, 687

\bibitem[Worthey \& Ottaviani(1997)]{worthott97}
Worthey, G., \& Ottaviani, D., 1994, \apjs, 111, 377

\bibitem[Wright(2006)]{wright06}
Wright, N., 2006, \pasp, 118, 1711

\bibitem[Yamanaka et al.(2009)]{yamanaka09}
Yamanaka, M. et al., 2009, \apj, 707, L118

\bibitem[York et al.(2000)]{york00} 
York, D., et al., 2000, \aj, 120, 1579

\bibitem[Yuan et al.(2010)]{yuan10} 
Yuan, R., et al., 2010, \apj, 715, 1338

\bibitem[Yungelson \& Livio(2000)]{yunglivio00} 
Yungelson, L., \& Livio, M., 2000, \apj, 528, 108

\bibitem[Zaritsky et al.(1994)]{z94}
Zaritsky, D., et al., 1994, \apj, 420, 87

\bibitem[Zhao, Gu, \& Gao(2011)]{zgg11} 
Zhao, Y., Gu, Q., \& Gao, Y., 2011, \aj, 141, 68

\end{thebibliography}
\end{document}